\documentclass[utf8]{frontiersSCNS} % for Physics and Applied Mathematics and Statistics articles

\usepackage{url,hyperref,lineno,microtype,subcaption}
\usepackage[onehalfspacing]{setspace}

\newcommand{\dnu}{$\Delta\nu$}
\newcommand{\numax}{$\nu_{\rm max}$}
\newcommand{\nuac}{$\nu_{\rm ac}$}
\newcommand{\logg}{$\log g$}
\newcommand{\teff}{$T_{\rm eff}$}
\newcommand{\dd}{{\hbox{d}}}

\def\keyFont{\fontsize{8}{11}\helveticabold }
\def\firstAuthorLast{Basu and Hekker} %use et al only if is more than 1 author
\def\Authors{Sarbani Basu\,$^{1,*}$ and Saskia Hekker\,$^{2,3}$}

\def\Address{$^{1}$Department of Astronomy, Yale University, New Haven, CT, USA \\
$^{2}$Max Planck Institute for Solar System Research, G\"ottingen, Germany \\
$^{3}$Stellar Astrophysics Centre, Department of Physics and Astronomy, Aarhus University, Aarhus, Denmark}
\def\corrAuthor{Sarbani Basu}

\def\corrEmail{sarbani.basu@yale.edu}

\begin{document}
\onecolumn
\firstpage{1}

\title[Structure and Dynamics of Red Giants]{Unveiling the Structure and Dynamics of Red Giants with Asteroseismology} 

\author[\firstAuthorLast ]{\Authors} %This field will be automatically populated
\address{} %This field will be automatically populated
\correspondance{} %This field will be automatically populated

\extraAuth{}% If there are more than 1 corresponding author, comment this line and uncomment the next one.
%\extraAuth{corresponding Author2 \\ Laboratory X2, Institute X2, Department X2, Organization X2, Street X2, City X2 , State XX2 (only USA, Canada and Australia), Zip Code2, X2 Country X2, email2@uni2.edu}

\maketitle

\begin{abstract}
The \textit{Kepler} mission observed many thousands of red giants. The long time series, some as long as the mission itself, have allowed us to study red giants with unprecedented detail. Given that red giants { are intrinsically luminous}, and hence can be observed from very large distances, knowing the properties of red giants, in particular ages, is of immense value for studies of the formation and evolution of the Galaxy, an endeavor known as ``Galactic archaeology''. In this article we review what we have learned about red giants using asteroseismic data. We start with the properties of the power spectrum and move on to internal structure and dynamics of these stars; we also touch upon unsolved issues in red-giant asteroseismology and the prospects of making further progress in understanding these stars.
\tiny
 \keyFont{ \section{Keywords:} Stellar oscillations, Fundamental parameters of stars, Stellar evolution, Stellar interiors, Stellar rotation} %All article types: you may provide up to 8 keywords; at least 5 are mandatory.
 
\end{abstract}

\section{Introduction}

\begin{figure*}
\centering
\begin{minipage}{0.49\linewidth}
    \includegraphics[width=\linewidth]{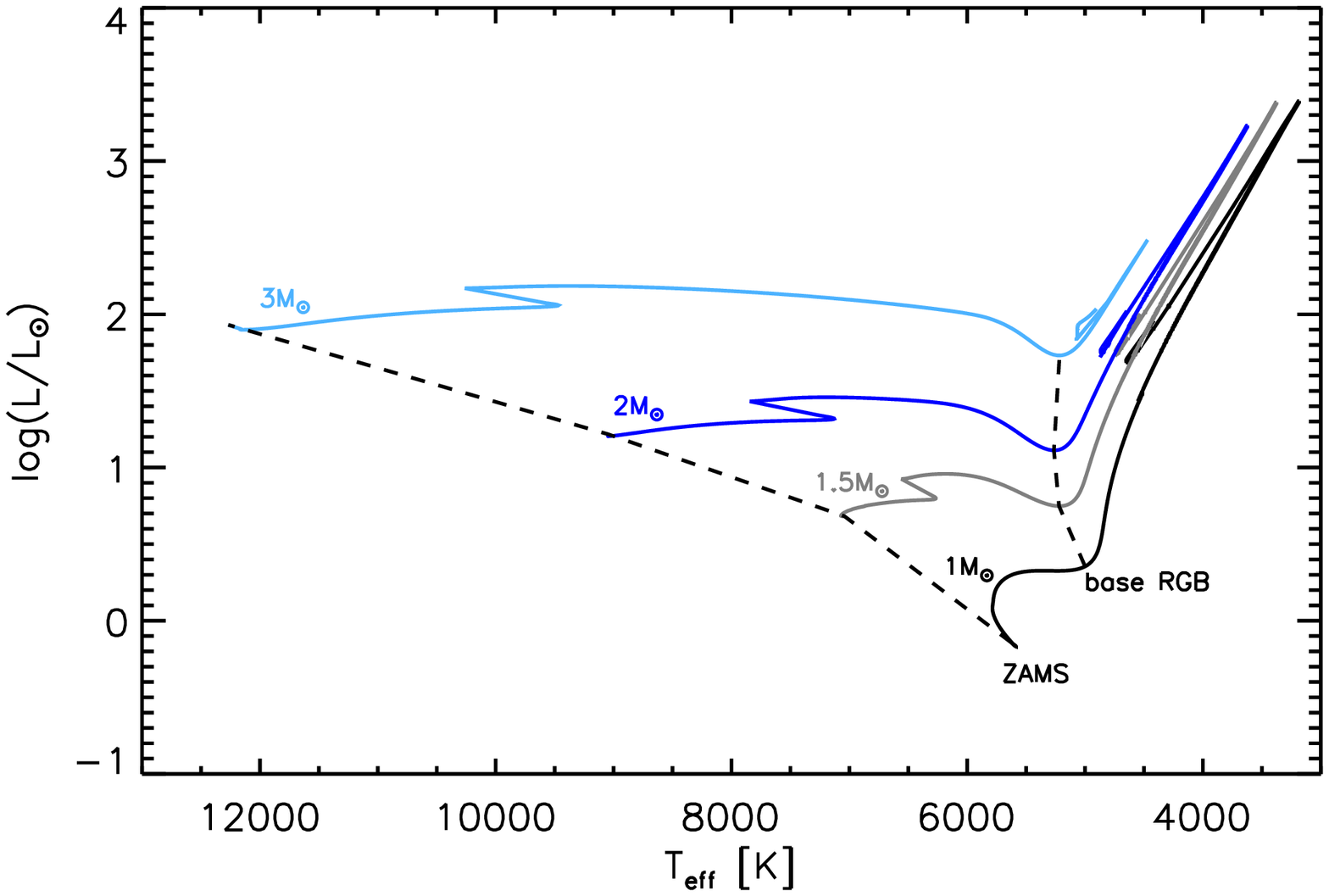}
\end{minipage}
\begin{minipage}{0.49\linewidth}
    \includegraphics[width=\linewidth]{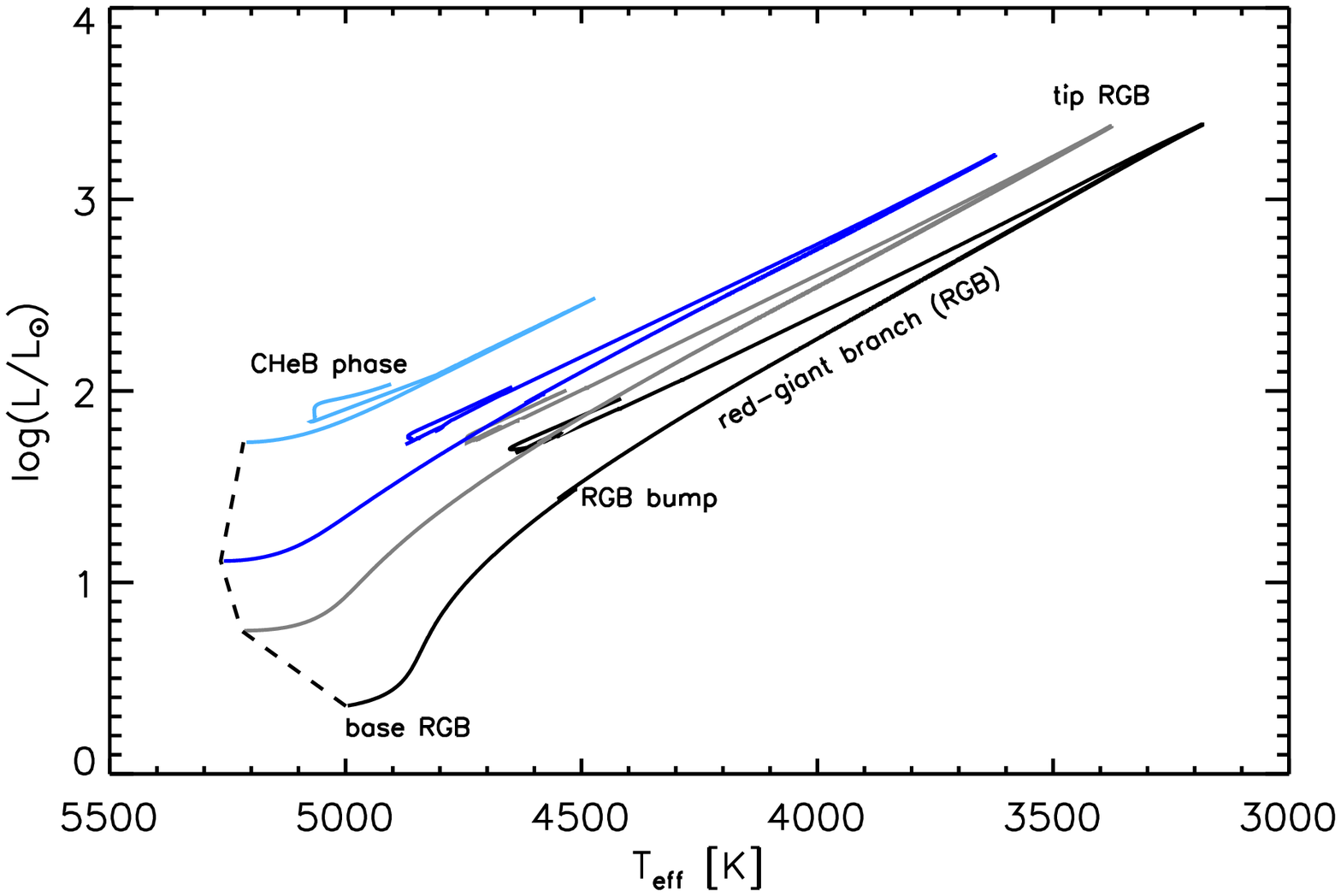}
\end{minipage}
	\caption{Left: HR Diagram with four evolutionary tracks of models of masses 1.0, 1.5, 2.0 and 3.0 M$_{\odot}$ in black, grey, blue and light blue respectively. The tracks are shown from the zero age main sequence (ZAMS) through the red-giant branch (RGB) and the core helium burning (CHeB) phase until the helium core mass fraction drops below $10^{-3}$. Right: the same tracks as in the left panel now starting from the base of the red-giant branch. The models were computed using MESA \citep[][and references therein]{paxton2019} version r-10398 and solar metallicity. {See Appendix A for the inlist of the MESA computations.}} 
    \label{fig:HRD}
\end{figure*}

Red-giant stars mark that stage of stellar evolution when a star has exhausted its central hydrogen, fusion occurs in a very thin shell around an inert helium core, and the envelope is cool enough that the convection zone encompasses most of the star making the stars almost fully convective {\citep[see, e.g.,][etc.]{salaris, kippenhahn}}. Being almost fully convective, the stars are constrained to take an almost vertical path on the Hertzsprung-Russell (HR) diagram (see Fig.~\ref{fig:HRD}). Along the red-giant branch (RGB) the stars exhibit a very narrow range of temperatures. The stars follow the vertical path{ , briefly disrupted by a short phase in which the star decreases its luminosity, i.e. the RGB bump,} until the onset of helium fusion in the core. For stars with masses below about 2~M$_{\odot}$ { at the tip of the RGB} the onset of helium fusion occurs in degenerate conditions. This is a fast process referred to as the helium-flash. For higher-mass stars, the onset of helium burning occurs in non-degenerate conditions. After the onset of helium burning in the core, the star reduces in luminosity and radius, and has a slightly higher temperature. High-metallicity low-mass stars in this stage settle into the red clump and higher-mass stars form the so-called secondary clump; low-metallicity stars form the horizontal branch. While these stars are fascinating on their own accord, their high luminosities make them visible from large distances, and thus useful tools in studying the Galaxy.

{Stellar oscillation frequencies can be derived from radial velocity measurements,
(e.g., from the Stellar Oscillation Network Group; \citealt{song}), as well as photometry from space with missions like {\it CoRot} \citep{corot} and {\it Kepler} \citep{kepler}}
The recent \textit{Kepler} space mission provided long ($\sim$ 4 years) of near-uninterrupted high-precision high-cadence photometric timeseries data. These data are of unprecedented quality and frequency resolution, and are a treasure trove for red-giant asteroseismology. This has led to many ground-breaking results published over the last decade. { These include the detection of g-mode period spacings \citep{beck2011}, the discovery that g-mode spacings can be used to distinguish inert-helium core red giants from core helium burning giants \citep{bedding2011,mosser2011}, the unequivocal detection of core rotation \citep{beck2012}, that red giant cores might actually slow down rather than spin up as they contract \citep{mosserspindown, gehan2018}, that precise estimates of red-giant surface gravities can be derived from contribution that granulation makes to the photometric time series \citep{kallinger2016}, etc. These data have also resulted in catalogues of stellar properties \citep[][etc.]{pinsonneault2014,pinsonneault2018} that are proving invaluable to the study of Galactic archaeology.} %\citep[e.g.][]{beck2011,bedding2011,beck2012,mosser2014,kallinger2016}. 
Many such  results are also highlighted in recent reviews by \citet{hekker2013, mosser2016, hekker2017}. In this review  we focus on the properties of the power spectra of red-giant stars and how they can be interpreted in terms of their  structure and dynamics.

\section{Properties of the power spectrum}
Red-giant stars are solar-like oscillators, i.e., their oscillations are stochastically excited and damped by near-surface turbulent convection { \citep[see e.g.,][]{goldreich1977a, goldreich1977b, balmforth1992}}. Unlike classical pulsators, many oscillation modes are excited, however the amplitudes of the modes are small. The frequencies of the modes are determined by the internal structure of the star, which makes asteroseismic inferences on stellar structure possible. Spherical harmonics are used to describe the angular dependence of the modes, with the degree $l$ being the number of nodal planes intersecting the surface, and the azimuthal order $m$ the number of nodes along the equator. In the radial direction, the models are described by the radial order $n$, which is the number of nodes in the radial direction. Conventionally, acoustic modes are denoted by positive values of $n$ and while gravity modes, i.e., modes with buoyancy as restoring force, are denoted by negative values of $n$. The azimuthal order comes into play only if rotation or magnetic fields (or both) break spherical symmetry. In the absence of these symmetry-breaking factors, there is a single mode with a given ($l,n$); rotation ``splits'' the modes into $2l+1$ components labelled by $m$.
For stars other than the Sun only low-degree oscillations  can be observed due to cancellation effects that are a result of observing stars as point sources. Typically one observes radial ($l=0$), dipole ($l=1$) and quadrupole ($l=2$) modes, though sometimes octupole ($l=3$) modes can be detected as well.
{ More details about the characteristics of oscillation frequencies and how they are related to internal
structure and dynamic are described in textbooks such as \citet{unno} and \citet{aerts}}.
Whether, and how many, rotationally split modes are observed depends on the inclination angle of the star 
{\citep{gizon2003}} assuming of course, that the frequency resolution is good enough. 
%In fact the three components of a rotationally split dipole mode can only be seen as as a triplet is a star is seen edge-on (i.e., along the equator); only one component can be seen if a star is seen pole-on, and two components at all other inclination angles. \Saskia{I don't think this is correct, shouldn't it be 3 components at intermediate angles, 2 at edge on and 1 at pole on. Furthermore, do we want to include here that the splitting needs to be resolved, i.e. if the star rotates too slow, we will also only see one mode due to the frequency resolution?} 
These rotationally split modes provide insight in the rotation rate from the size of the splitting and on the inclination angle from the relative amplitudes of the split components.

\begin{figure*}
\centering
\begin{minipage}{\linewidth}
    \centerline{\includegraphics[width=0.75\linewidth]{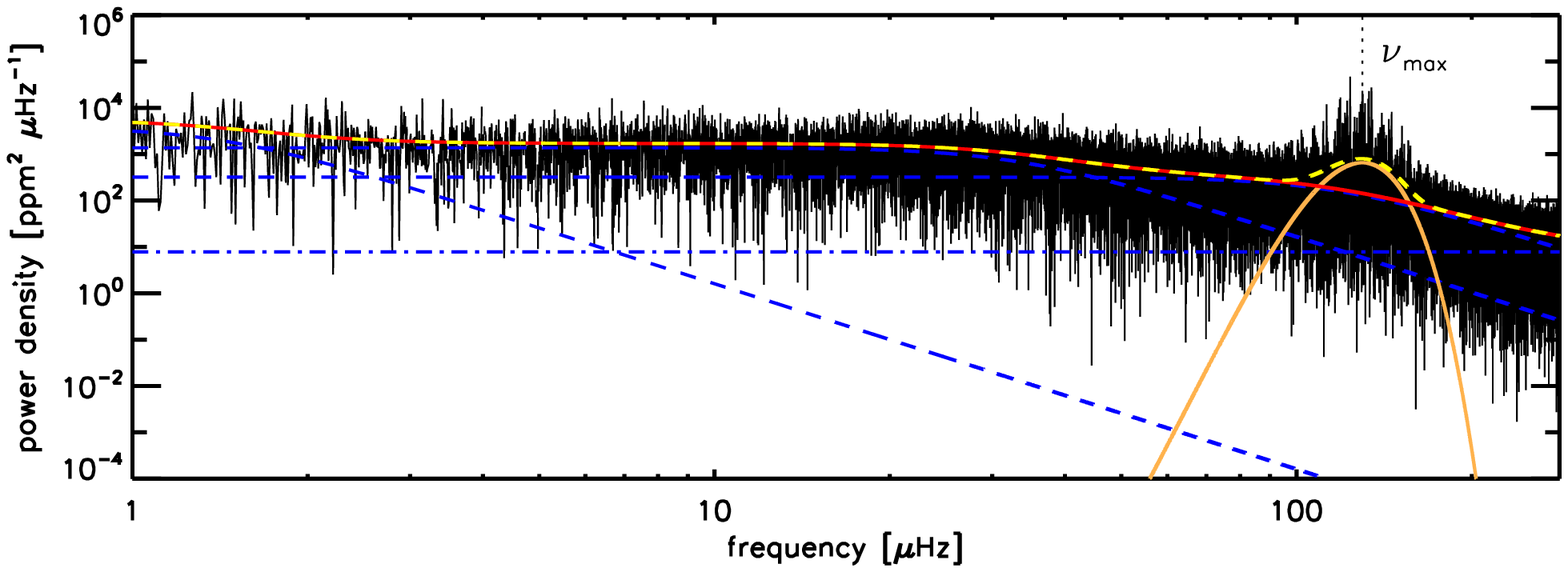}}
\end{minipage}
\begin{minipage}{\linewidth}
    \centerline{\includegraphics[width=0.75\linewidth]{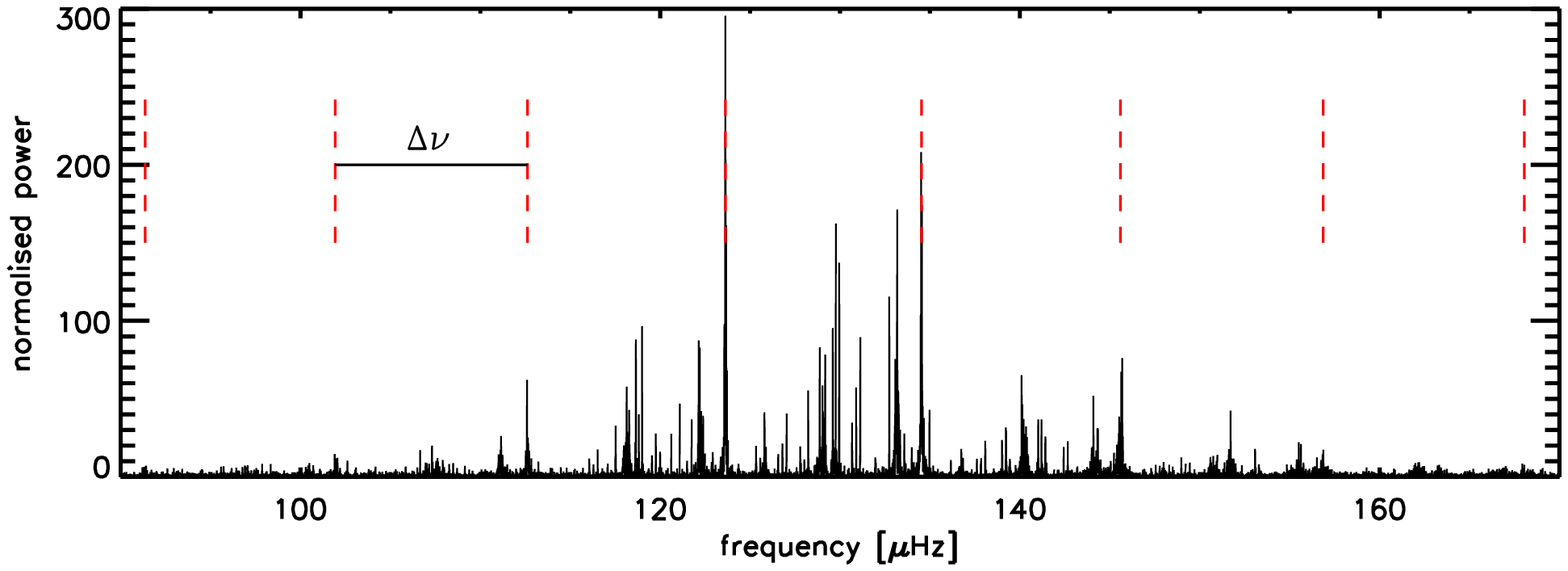}}
\end{minipage}
	\caption{Top: power density spectrum of KIC~6144777 with the data shown in black. The granulation components are indicated with blue dashed lines and the white noise with the blue dashed-dotted line. The sum of the granulation components and noise comprise the background fit, which is shown with the red solid line. The Gaussian fit to the power excess is shown in orange with the frequency of maximum oscillation power ($\nu_{\rm max}$) indicated by the vertical dotted line. The complete fit (background + oscillation power excess) is shown with the yellow dashed line. Bottom: the power spectrum after the background correction. The regular pattern of radial modes is indicated with the vertical red dashed lines. The distance in frequency between these lines is the large frequency separation $\Delta\nu$.} 
    \label{fig:PDS}
\end{figure*}

We show a power spectrum of a typical red giant in Fig.~\ref{fig:PDS}.
The power density spectrum of red-giant stars, as well as other solar-like oscillations such as low-mass main-sequence stars and subgiants, is dominated by granulation --the observable part of the near-surface turbulent convection-- as a frequency-dependent signal and is commonly modelled as red noise.
The power spectrum is modulated by a broad bell-shaped envelope whose maximum is at a frequency often called $\nu_{\rm max}$, the frequency of maximum oscillation power. The value of $\nu_{\rm max}$ provides a direct rough indication of the size of the star, i.e. it takes longer for the waves to travel through a larger star and hence stars with larger radii have oscillations occurring at lower values of $\nu_{\rm max}$. \citet{brown1991,KB1995} argued that $\nu_{\rm max} \propto \nu_{\rm ac}$, where \nuac\ is the acoustic cutoff frequency i.e., the frequency beyond which acoustic modes are no longer trapped in the star and behave as traveling waves. Under the assumption of an isothermal atmosphere, one can show that $\nu_{\rm ac}$ is proportional to $g/\sqrt{T_{\rm eff}} \propto M/(R^2\sqrt{T_{\rm eff}})$, where $g, M, R, T_{\rm eff}$ are surface gravity, mass, radius and effective temperature, respectively. 
Both the shape of the envelope and the value of $\nu_{\rm max}$ are determined by the excitation and damping of the oscillation modes. There is as yet no theoretical understanding as to why \numax\ is proportional to \nuac.

Another property of the power spectra of solar-like oscillators is the picket-fence, or comb-like, pattern of low degree modes. The modes of a given degree $l$ and consecutive radial orders $n$ are approximately equidistant in frequency, and the separation is called the   large frequency separation, $\Delta\nu$.   \citet{ulrich1986,jcd1988} showed that the large frequency separation scales approximately with the square root of the mean density ($\bar{\rho}$) of the star: $(\Delta\nu/\Delta_{\nu_\odot})\sim \sqrt{(\bar{\rho}/\bar{\rho_\odot}}$.  The expression is not exact and has well known deviations that are a function of temperature \citep{whiteetal2011}, metallicity \citep{guggenberger2016} and at the red-giant end the deviation also depends on mass \citep{guggenberger2017}. For an extensive overview of ways to mitigate these deviations see \citet{hekker2020} and references therein. However, despite these deviations, the $\Delta\nu$ scaling relation, along with the relation for \numax, have been very useful in getting initial mass and radius estimates of all types of stars \citep{chaplin2010, pinsonneault2014,pinsonneault2018}. As shown by \citet{ong2019}, we now understand why these deviations occur, and have derived ways of calculating \dnu\ for models that give much better approximations to \dnu\ , without resorting to calculating the frequencies.

\begin{figure*}
\centering
\begin{minipage}{\linewidth}
\centering 
    \includegraphics[width=0.85\linewidth]{{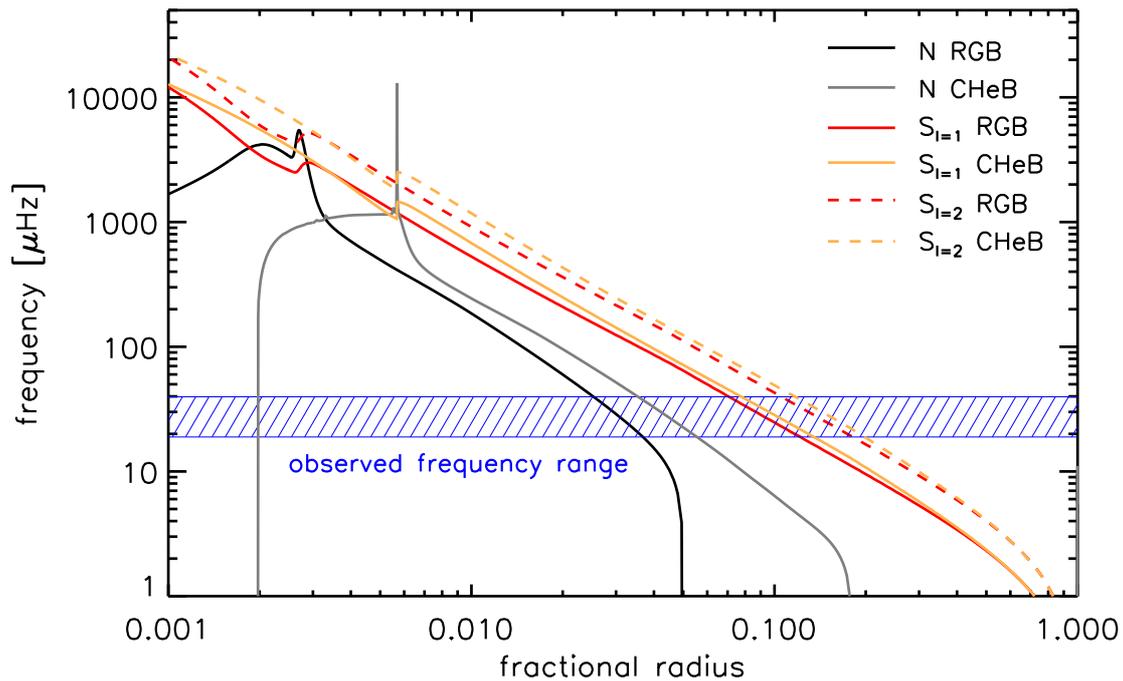}}
\end{minipage}
	\caption{Propagation diagram of a 1~M$_{\odot}$ model of an ascending the red-giant branch (RGB) star and a core helium burning (CHeB) star on the same evolutionary track. For the RGB model, the Brunt-V\"ais\"al\"a frequency $N$ is shown in black and the Lamb frequency $S$ for dipole ($l=1$) and quadrupole ($l=2$) modes are shown with the red solid and dashed line, respectively. The Brunt-V\"ais\"al\"a frequency and Lamb frequencies for the core helium burning (CHeB) model are indicated in gray and orange. The region around the $\nu_{\rm max}$ value of both models is indicated by the horizontal blue bar.} 
    \label{fig:PROP}
\end{figure*}

\begin{figure*}
\centering
\begin{minipage}{\linewidth}
    \centerline{\includegraphics[width=0.85\linewidth]{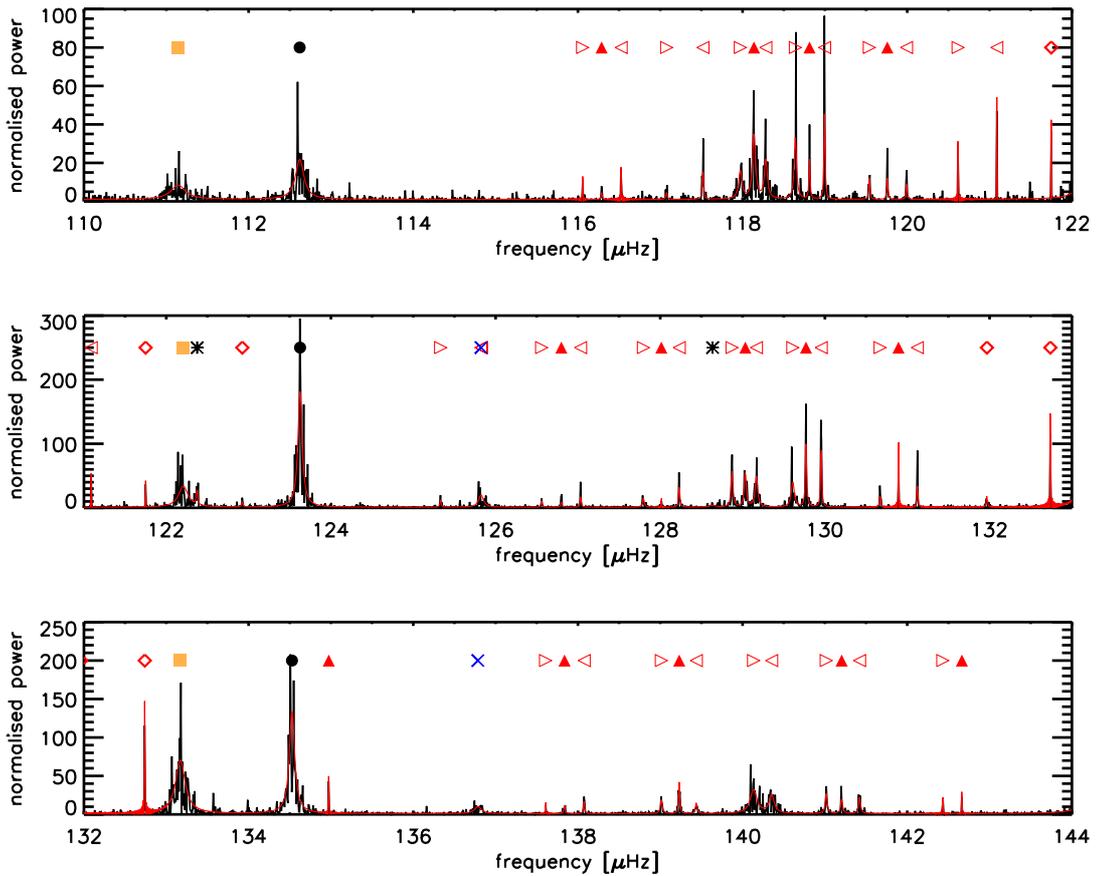}}
\end{minipage}
	\caption{A close look into the region of the three radial orders closest to \numax\ in the power density spectrum of KIC~6144777. Each row focuses on one radial mode and the adjacent $l=1$ and $l=2$ modes. The fit to the data as computed with TACO (Hekker et al, in prep) is indicated in red. The radial, dipole, quadrupole and {octupole} modes are indicated with black dots, red triangles, orange squares and blue crosses respectively. Note that there are multiple dipole modes associated with each radial mode --- these are the mixed modes. Additionally, the dipole modes show rotational splitting, where the $m=-1$ and $m=+1$ modes are indicated with right and left pointing open triangles respectively. Dipole modes for which the azimuthal order could not (yet) be identified are indicated with red diamonds. Modes for which no identification could be obtained are indicated with black asterisks.} 
    \label{fig:PDSfit}
\end{figure*}

Solar-like oscillation modes can be divided into acoustic modes (i.e., sound waves with pressure as the restoring force) or gravity modes (with buoyancy as the restoring force). The acoustic modes are the ones that are equidistant in frequency and separated by \dnu; gravity modes on the other hand are equidistant in period. Whether a mode is acoustic or gravity depends on two frequencies: the Lamb frequency $S_l$ which depends on the sound-speed profile of the star as well as the degree of a mode; and the buoyancy or Brunt-V\"ais\"al\"a frequency $N$ that is related to the convective stability criterion and is imaginary in convection zones. Acoustic modes have frequency $\omega$ such that $\omega^2 > S_l^2$ and $\omega^2 > N^2$, and gravity modes occur in the region where $\omega^2 < N^2$ and $\omega^2 < S_l^2$. For main-sequence solar-like oscillators, the acoustic modes and gravity modes are trapped in well separated regions of a star and the observed modes are purely acoustic modes. However, in
 evolved solar-like oscillators the non-radial ($l>0$) oscillation modes are not necessarily pure acoustic modes. 
As a star evolves, the density of the core increases leading to an increase of the buoyancy frequency values, while at the same time the outer layers of the star expand leading to a decrease of \numax. This can result in a coupling of the acoustic and
 gravity mode cavities.
 In Fig.~\ref{fig:PROP} we show a so-called ``propagation diagram'' for red-giant models. This diagram shows the Lamb frequency for  $l=1$ and $l=2$ modes along with the buoyancy frequency.  We can see that
 around \numax\ for these models, the radial distance between the cavity for acoustic $l=1$ modes and for gravity modes is small, allowing g modes (gravity modes) to couple with p modes (acoustic modes). The ``mixed'' modes
  have gravity-mode characteristics in the deep interior, and acoustic mode characteristics in the outer layers. In red giants, multiple g-modes can couple with a single p mode leading to multiple mixed modes per acoustic radial order. The presence of mixed modes causes multiple peaks for a given $l=1$ p mode. We show
  this in Fig.~\ref{fig:PDSfit}. The mixed modes of degree $l=1$ (dipole modes) are most evident in red-giant power spectra. Coupling is weaker for the $l=2$ (quadrupole) modes because of the larger evanescent zone between the Lamb and Brunt-V\"ais\"al\"a frequency (see Fig.~\ref{fig:PROP}), as a result only the most p-dominated quadrupole modes are observed easily.
Additionally, the $l=2$ g-dominated modes would be harder to resolve since the frequency differences between the mode p-dominated and the neighbouring g-dominated modes is often unresolved given the frequency resolution and width of the modes. 
{More about the nature of the modes can be found in \citet[][etc.]{unno, aerts}}

The mixed dipole modes appear at frequencies that are equidistant in period ($\Delta\Pi$) as dictated by the gravity-part of the mode, modulated by the strength of the coupling to the pressure part of the mode {\citep[e.g.][]{mosser2012}}. In other words, close to the nominal pressure mode the coupling is strongest and the observed period spacing deviates from the underlying period spacing of the gravity part of the mode, while further away from the nominal pressure mode the period spacing approaches the underlying value. This underlying period spacing acts as a direct probe of the core of red-giant stars. 
  
\begin{figure}[htb]
\centering
\begin{minipage}{0.40\linewidth}
\caption{The {\'e}chelle diagram of the modes of KIC~6144777. As in Fig.~\ref{fig:PDSfit}, the black dots are radial modes, the red symbols are dipole modes with the $m=0$ indicated by filled triangles, while the $m=-1$ and $m=+1$ modes are indicated with right and left pointing open triangles, respectively, the dipole modes with unidentified azimuthal orders are indicated with red diamonds. The orange squares are the quadrupole mode and the blue crosses the { octupole} ($l=3$) modes. Modes that could not be identified are indicated by black asterisks. Note that the radial modes line up vertically, but not perfectly, and there is a curvature (see text for more details).}
\label{fig:ECHELLE}
\end{minipage}
\begin{minipage}{0.59\linewidth}
\includegraphics[width=\linewidth]{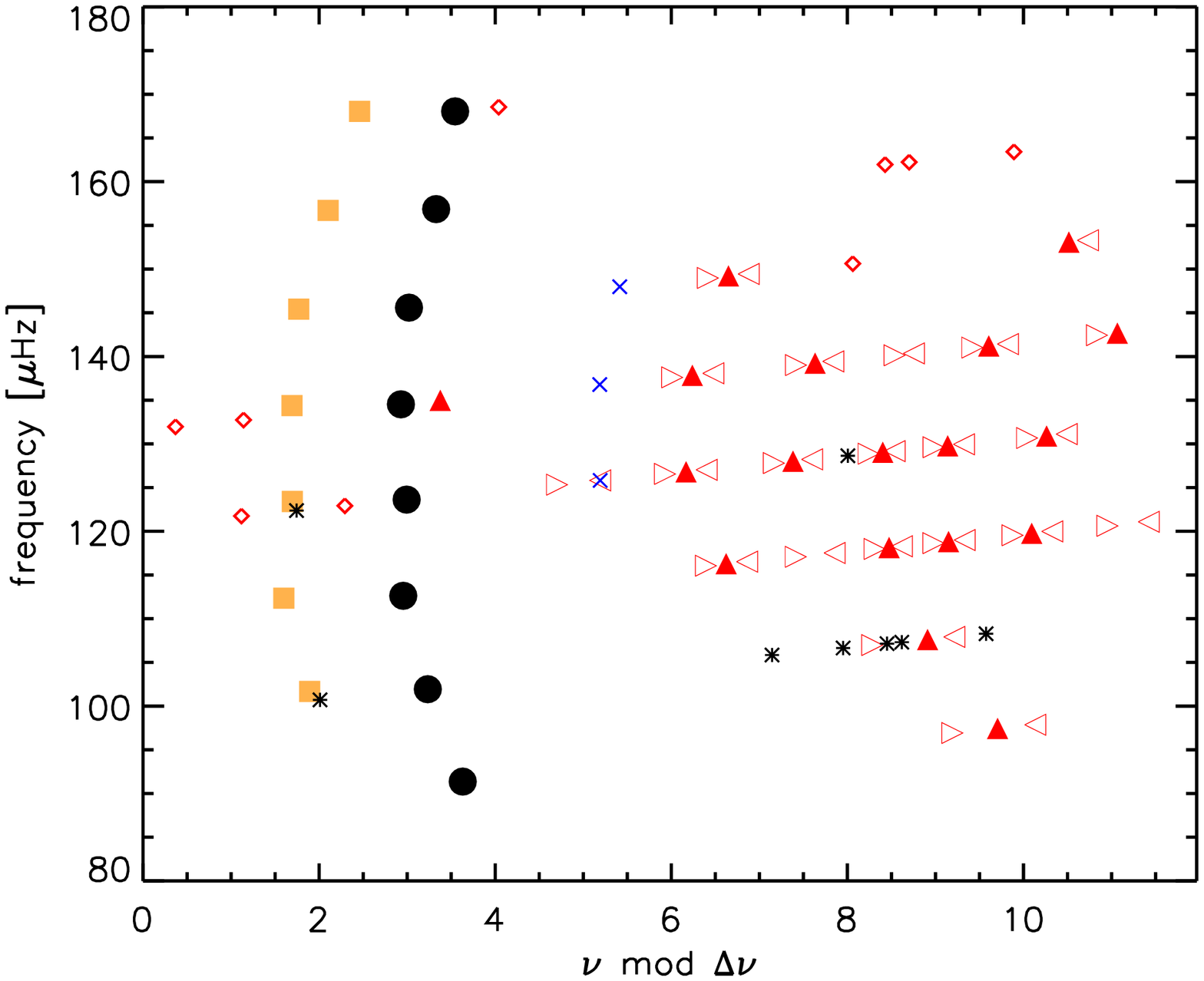}
\end{minipage}
\end{figure}

In reality, mode-frequencies do not follow a strict separation in frequency (or period for g modes), i.e., the frequency/period separation between adjacent modes differs from the average value of $\Delta\nu$ or $\Delta\Pi$, although these deviations are small. These deviations are a result of the fact that stars are not homogeneous balls of
gas, but are stratified, and this leaves a `curvature' in the frequency spacings that is easily seen in an {\'e}chelle diagram of the frequencies (see Fig.~\ref{fig:ECHELLE}).
 Abrupt changes, so-called glitches, leave an oscillatory pattern in the frequency separations, and the deeper inside the star
 the location of the glitch occurs, the shorter the wavelength of the pattern {\citep{dog1990}}. The base of the convection zone, and the helium second ionisation zone are places where an acoustic glitch occurs. In evolved stars, the hydrogen-burning shell can leave a
 glitch in the buoyancy frequency which affects period spacings. These glitches provide a direct probe of the location and nature of the change in the stellar structure. We refer reader to \citet{mybook} for details on the behaviour of acoustic, gravity and mixed modes in different types of solar oscillators, as well as the effect of glitches and how they may be used to study stellar structure.

\vspace{0.75 true cm}

\section{Differentiating red-giant and red-clump stars}

One of the important results obtained with {\it Kepler} data was the realization that asteroseismic data can be used to differentiate RGB stars from core-helium burning red-clump stars. Red-clump stars are important because they are standard candles since their luminosity depends very weakly on mass and metallicity \citep{girardi2016}. These stars can be used as distance tracers out to $\sim 10$~Kpc { \citep[e.g.][]{bovy, mathur02016, ting}}. Currently, the Gaia mission is providing parallaxes for billions of stars, however, beyond about 3~Kpc Gaia uncertainties can become larger than those provided by standard candles such as red-clump stars \citep[e.g.][]{mints2018}. In fact red-clump stars have revealed zero-point errors in Gaia data and have aided in showing how these errors can be corrected \citep{guy, saniya}.  However, because red-clump stars occur in a  \teff\ and \logg\ range where ascending branch red-giant stars are also found, it is almost impossible to pick them out among field red giants; in clusters on the other hand, the over-density of points in the color-magnitude diagram (CMD) at the characteristic \logg\ facilitates detection.

\begin{figure}[htb]
\centering
\begin{minipage}{0.40\linewidth}
\caption{Observed mixed-mode spacings for a sample of red giants observed by {\it Kepler}. Data are from \citet{bedding}.}
\label{fig:DP}
\end{minipage}
\begin{minipage}{0.59\linewidth}
\centering
\includegraphics[width=0.95\linewidth]{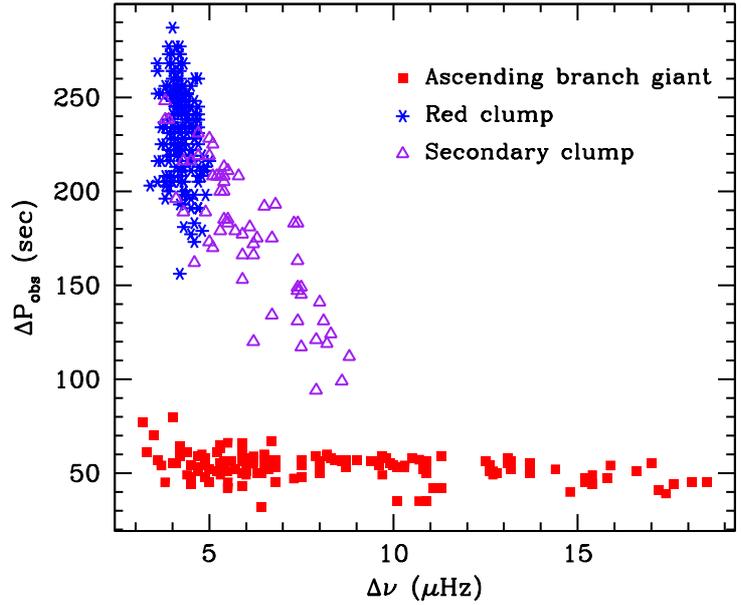}
\end{minipage}
\end{figure}

\citet{bedding} showed that asteroseismic data can distinguish between the two types of giants, and the key discriminant is the period spacing of the dipole { model} in the two types of stars (see Fig.~\ref{fig:DP}) with red-clump stars having higher period spacings than red giants. Secondary clump stars, i.e., stars massive enough to initiate helium burning before their cores become degenerate have intermediate values. It is actually quite easy to understand the result, and in fact, it should have been anticipated. Asymptotically, the period spacing ($\Delta\Pi_0$) of the g-mode oscillations can be expressed as
$\Delta\Pi_0=2\pi^2(\int_{r1}^{r2} N(r) \dd r/r)^{-1}$, where $r_1$ and $r_2$ are the boundaries of the radiative zone; for modes of a given $l$, the spacing ($\Delta\Pi_l$) is then $\Delta\Pi_0/\sqrt{l(l+1)}$ {\citep{Aizenman1977, tassoul1980}}. The key to understand the difference in $\Delta\Pi$\footnote{We drop the degree identification here as we focus only on dipole mixed modes.} for RGB and CHeB stars is in the integral over the buoyancy frequency. As can be seen in Fig.~\ref{fig:PROP}, the core convection zone in RC stars means that the integral is smaller, and hence $\Delta\Pi$ is larger. The utility of the period spacing has led to the development of automated methods of determining $\Delta\Pi$ values \citep[e.g.,][]{vrard2016} and from that the evolutionary phases of observed red-giant stars. 

\begin{figure}[htb]
\centering
\begin{minipage}{0.40\linewidth}
\centering
\caption{The asteroseismic phase function $\epsilon$ at \numax\ as a function of $\Delta\nu$ for stars observed with {\it Kepler}. The pink crosses are RGB and RC stars from \citet{kallinger2012}, with the red-clump stars forming a distinct group around a \dnu\ of 4~$\mu$Hz. 
KAGES refers to the exoplanet-host sample of \citet{kages, sva2015}, and LEGACY to the sample of main-sequence stars in \citet{lund, sva2017}. \citet{thierry} has a mix of main-sequence and subgiant stars. {NGC~6791 refers to the sample analyzed by \citet{mckeever}, and ``TESS subgiants'' are a set of sungiants in the TESS southern continuous viewing zone that show clear oscillations.} Image courtesy of Joel Ong.}
\label{fig:EPS}
\end{minipage}
\begin{minipage}{0.55\linewidth}
\centering
\includegraphics[width=0.93\linewidth]{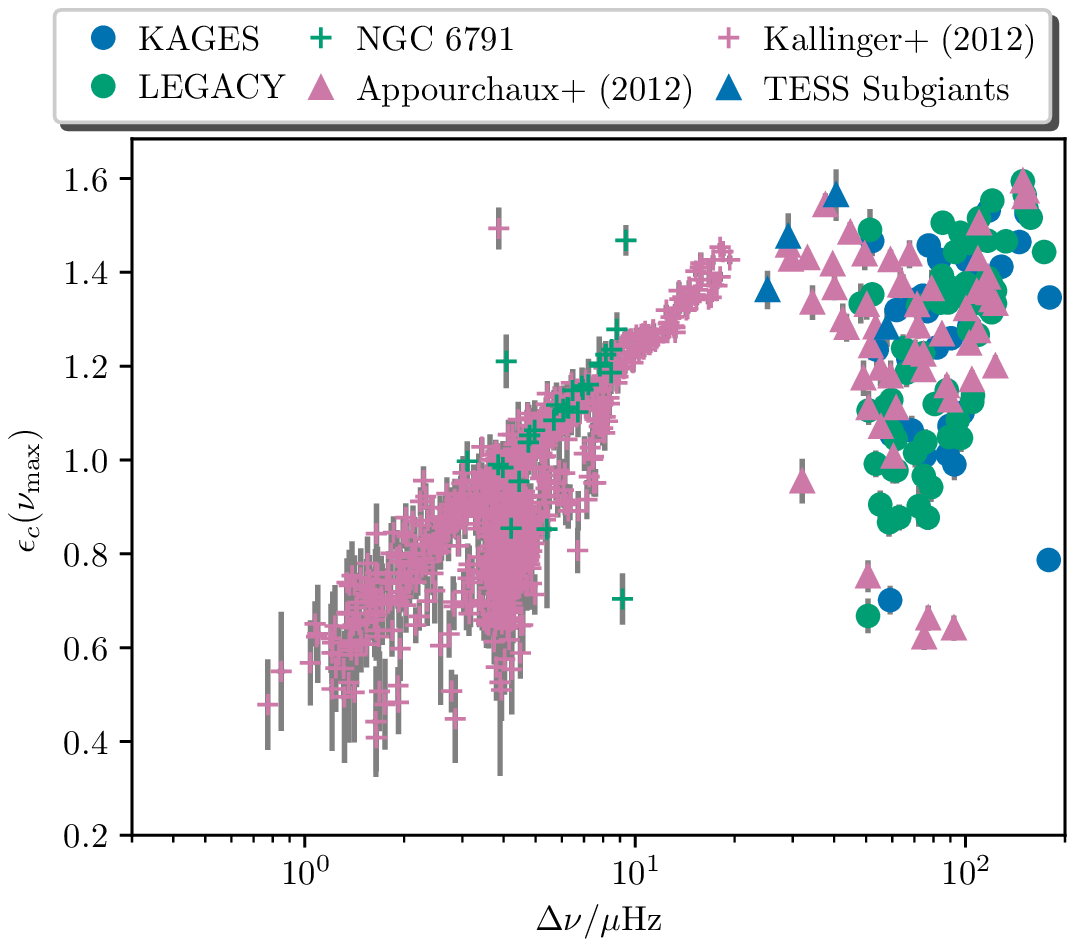}
\end{minipage}
\end{figure}

Deriving a reliable value of $\Delta\Pi$ requires a power spectrum that has a reasonably high signal-to-noise, which is not always the case. \citet{kallinger2012} showed that the asteroseismic phase function $\epsilon$ can help. Asymptotically, frequencies follow the relation $\nu\simeq\Delta\nu(n+l/2+\epsilon),$ where $\epsilon$ is the `phase function'. The phase function can be a function of frequency (due to the curvature mentioned earlier), though the value of $\epsilon$ around \numax\ can be defined in a straight forward manner. \citet{kallinger2012} showed that in a plot of $\epsilon$ against \dnu\ or \numax, red-clump stars stand out. There has been some attempts to explain why the phase function behaves different in red-clump stars. \citet{JCD2014}\ showed that the differences between the phase functions of red-giant and red-clump stars arise from differences in the thermodynamic state of their convective envelopes. \citet{epsilon} subsequently showed that this sensitivity to evolutionary stage arises from differences in the local frequency derivative of the underlying phase function which has a large contribution from the interior of the star too, and as a result this can be used to classify other types of stars as well;
 interestingly all the RGBs follow a much more well-defined sequence than stars in other states of evolution (Fig.~\ref{fig:EPS}).

The importance of red-clump stars to Galactic archaeology has led to the development of a number of automated pipelines that look at different seismic properties in order to classify the stars. These include using a method that uses the measured period spacings \citep{mosser2014,mosser2015}, the morphology of the power spectrum \citep{elsworth2017}, a grid-based method using the ratio
$\Delta\nu/\nu_{\rm max}$ \citep{hekker2017pipeline}, use of the phase factor $\epsilon$ \citep{kallinger2012}, as well as some machine learning based methods based on the power spectrum \citep{hon}, and  based directly on the timeseries data (Kuszlewicz et al. submitted). None of the techniques is perfect, and a comparison of some of the methods can be found in \citet{elsworth2019}, where the consensus evolutionary stage of the sample of  red giants in \citet{pinsonneault2018} has been presented.

\vspace{0.75 true cm}

\section{Structure and dynamics}
\label{sec:strucdyn}

\begin{figure*}
\centering
\begin{minipage}{\linewidth}
    \centerline{\includegraphics[width=0.75\linewidth]{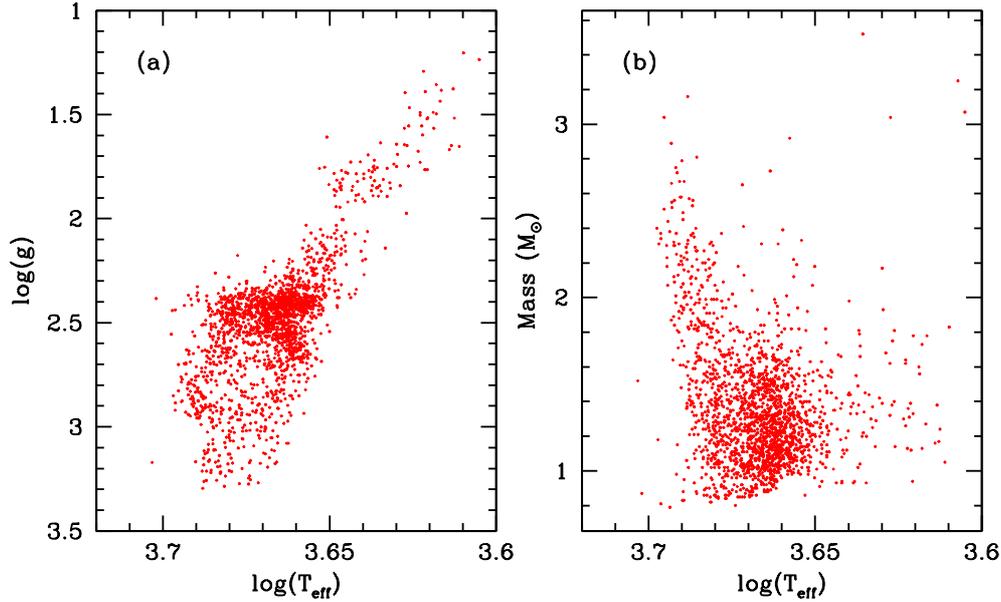}}
\end{minipage}
	\caption{Asteroseismically derived surface gravity (panel a) and mass (panel b) of the sample of red giants in \citet{pinsonneault2014}. The over-density of points in panel~(a) around $\log g \sim 2.5$ is caused by the presence of red-clump stars. There is a corresponding over-density in masses, though here the overdensity is confined to low masses; higher-mass stars do not form red-clump stars.} 
    \label{fig:APOKASC}
\end{figure*}

Most of the asteroseismic characterization of red giants that has been done thus far has involved determining the global properties of the stars --- surface gravity, masses, radii and ages --- with the ultimate aim of using these to study the chemo-dynamical evolution of the Galaxy. \citet{pinsonneault2014,pinsonneault2018} released a catalogue of red-giant properties based on asteroseismic data combined with spectroscopy. The uncertainties in the asteroseismic parameters were low enough that the red-clump stood out clearly in a Kiel diagram (see Fig.~\ref{fig:APOKASC}). These have been used extensively in studying the evolution of the Galaxy \citep[see, e,g,][etc.]{nidever2014, valentini2019, lian2020, spitoni2020}, and have also been used as training sets to for machine-learning based methods of deriving red-giant properties from spectra alone \citep[e.g.,][]{martig2016}.
The stellar properties derived in \citet{pinsonneault2014} were determined from \dnu, \numax, \teff\ and metallicity by searching among pre-computed grids of models. A different approach was taken by 
 \citet{pinsonneault2018}; in this paper masses and radii were determined using the scaling relations for \numax\ and \dnu, after correcting the \dnu\ scaling relation empirically using asteroseismic data of stars in clusters. This work was prompted in part by the result that the use of scaling relations yield masses for halo stars that are well above reasonable values for old stellar populations \citep{epstein2014}. One of the interesting discoveries made with such analyses is that there are high-mass stars that have
 a high [$\alpha$/Fe] ratio \citep{martig2015, chiappini2015}. This is interesting because a high [$\alpha$/Fe] ratio has been believed to be a hallmark of old, and hence, low-mass stars. \citet{hekker2019} and references therein proposed that the origin of these `$\alpha$-rich young' stars are main sequence merger remnants.

\vspace{0.75 true cm}

\subsection{Structure}
 Global asteroseismic parameters are not sufficient to determine the internal structure of red giants. Information about the structure is however, imprinted on the individual oscillations.
While there have been concerted efforts in modelling the individual frequencies of main-sequence stars \citep[e.g.,][]{sva2015, sva2017}, there has not been such efforts in the case of red giants, even though there are thousands of red giants with high-signal-to-noise oscillation power spectra. One of the main reasons is that the extraction of this wealth of information present in the data has caused difficulties that have been slow to overcome. Some codes have now been developed to extract and identify these modes \citep[e.g.][and Hekker et al. in prep.]{corsaro2014,garcias2018,kallinger2019}. Another main reason for the lack of results based on individual frequencies is the incredibly complicated frequency structure with all the mixed modes. The frequencies of mixed modes depend on details of the buoyancy-frequency profile, which in turn depends on uncertain internal processes such as convective overshoot and mixing; this is a challenge in modelling. Fortunately, the uncertainties of the modeling outcome will be reduced by accurate data on mixed modes. 

\begin{figure}[htb]
\centering
\begin{minipage}{0.45\linewidth}
\centering
\caption{The effects of the \citet{bg14} surface-term correction on $l=1$ modes of red-giant models. Note that the most p-dominated mode shifts enough to confuse its frequency with the more g-dominated mode at lower frequency. Data are from \citet{balletal2018}.}
\label{fig:BADCORR}
\end{minipage}
\begin{minipage}{0.50\linewidth}
\centering
\includegraphics[width=0.90\linewidth]{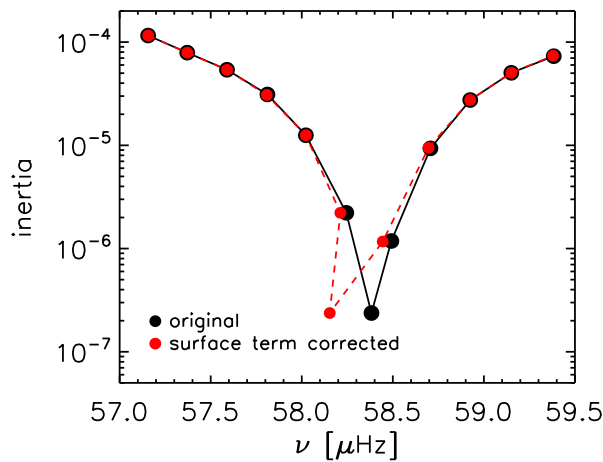}
\end{minipage}
\end{figure}

A more practical challenge in modelling the frequencies of red giants is the issue of the so-called `surface term'.
The surface term is a frequency-dependent frequency offset between models and stars caused by our inability to model the near-surface layers of a star properly, the main culprit being the approximations used to model convection. The way this is taken care of in main-sequence stars is to subtract out a slowly varying function of frequency from the models. There are many ways of doing this \citep[see e.g.][etc.]{kbj, bg14, sonoi}, though currently the most popular model is that of \citet{bg14}, who, based on work by \citet{dog1990} claimed that the surface term produces a frequency difference that can be expressed as
{
\begin{equation}
\delta\nu_{nl} =\nu_{nl}^{\rm obs}-\nu_{nl}^{\rm model}=\frac{1}{I_{nl}}\left[ a\left(\frac{\nu_{nl}}{\nu_{\rm ac}}\right)^{-1}
+b\left(\frac{\nu_{nl}}{\nu_{\rm ac}}\right)^3\right],
\label{eq:bl}
\end{equation}
}
where $\delta\nu_{nl}$ is the difference in frequency $\nu_{nl}$ for a mode of degree $l$ and order $n$
between a star and its model,  $I_{nl}$ is
the inertia of the mode, and $\nu_{\rm ac}$ is the acoustic cut-off frequency.
The coefficients $a$ and $b$ can be determined through a generalized linear least-squares fit. This form appears to work quite well for main-sequence stars and subgiants, and even for radial modes of red giants \citep{joey2015,balletal2018}. However, the 
$l=1$ modes of red giants pose a challenge. Mixed modes, particularly the g-dominated ones, have large inertia and hence the surface effect is small, as a result they do not change much when a surface-term correction is applied. Consequently the usual surface-effect corrections can be large enough to put the non-radial mixed modes in red giants out of order, which is unphysical \citep[see][and Fig.~\ref{fig:BADCORR}]{balletal2018}. \citet{balletal2018} proposed a method to modify the structure of the models to suppress g modes in the cores of  stellar models, so that they have only pure p modes on which to apply the surface-term correction. The authors showed that this did work for three double-lined eclipsing binaries,  KIC~8410637, KIC~9540226, and KIC~5640750. Note that suppressing the g-modes means that information from the core is neglected, but this provides a possible way to begin exploiting the data, and allows us to determine the mass and age of the star quite precisely. There are efforts under way (Ong and Basu, in preparation) of calculating pure p- and g-mode frequencies of stars without resorting to modifying the structure of the models. There are also efforts underway to determine red-giant surface-term corrections from simulations of stellar convection \cite[see e.g.,][]{newsurface}.

\begin{figure}[htb]
\centering
\begin{minipage}{0.45\linewidth}
\centering
\caption{The cyan curves mark the individual probability density function for age for a sample of red giants in
NGC 6791. The red curve is the joint probability assuming that all stars in a cluster have the same age. Note the precision to which the age of the cluster can be determined. Image courtesy of Jean McKeever.}
\label{fig:AGE}
\end{minipage}
\begin{minipage}{0.50\linewidth}
\centering
\includegraphics[width=0.90\linewidth]{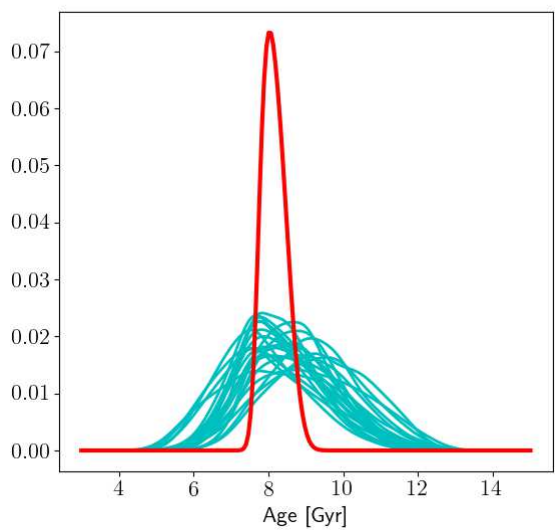}
\end{minipage}
\end{figure}

Masses and ages of red giants can be determined quite precisely from their radial and quadrupole modes, where the surface term behaves well. This was exploited by \citet{mckeever} to determine the age and initial helium abundance of the cluster NGC~6791 using cluster red giants. NGC~6791 is a rather strange cluster --- it is old, and yet very metal rich. Although the ages, or for that matter the helium abundance, of individual stars could not be determined to a high precision, using the fact that all cluster stars have the same age and initial metallicity allowed them to obtain a precise age of $8.2 \pm 0.3$~Gyr (see Fig.~\ref{fig:AGE}) and an initial helium abundance of $ Y_0 = 0.297\pm0.003$. This is to date, the most precise age and helium estimate for this cluster.

\vspace{0.75 true cm}

\subsection{Dynamics}
Unlike the case of red-giant structure, a lot more attention has been paid to the internal rotation of red giants.
\citet{beck2012} showed unequivocally that rotational splittings of red-giant modes can be observed and measured. This opened up the possibility of determining the rotation of RGB cores since mixed modes are most sensitive to conditions in the core. \citet{beck2012} and \citet{beck2012a} showed that the cores of red giants rotate
{ten times} faster than the surface, as is expected from the principle of conservation of angular momentum --- as stars evolve, the core contracts while the envelope expands, and the contracting core rotates faster to conserve angular momentum. 

Clearly, interpreting the rotational splittings require the knowledge of how the splittings depend on internal rotation. \citet{goupil2013} showed how the contrast between core and envelope rotation are encoded in the frequency splittings. They showed that the splittings have a linear dependence on the contribution of the core to the total mode inertia and that the slope { of this linear relation} is related to the ratio of the average envelope rotation to that of the core. The linear theory of rotational splittings predicts symmetric splittings around the $m=0$ mode for rotation that does not depend on latitude. However, this is not the case for strongly couple mixed modes. { Such asymmetries were first observed by \citet{beck2014}}. \citet{degeneracy} showed how the splittings might be calculated under conditions of strong coupling. Their work implied that it is important to  obtain a good model of the star being studied in order to obtain the eigenfunctions needed to interpret the data.

Rotational splittings only indicate an average rotation rate, and one has to invert the rotational splittings to determine what the real rotation rate is. \citet{otto} inverted the splittings of KIC~7341231, a star very close to the base of the red-giant branch, and found that the core of the star is rotating at least five times faster than the envelope. This result is interesting because while it shows that the core of stars spin up as they evolve, the spin up is not as fast as simple models of angular momentum transport in stars would suggest. \citet{sebastien_six} examined six more stars, all evolved subgiants and red-giants low on the red-giant branch. They obtained estimates of their core rotation rates and upper limits for the rotation in their convective envelopes. The results by \citet{sebastien_six} confirmed that the contrast between the rotation rates of the core and the envelope increases along the subgiant branch. They also find evidence of an abrupt change in the radial rotation profile in two of the stars, with the hydrogen-burning shell being the likely location of this change.
The evidence for a an abrupt change in rotation rate causing a shear layer was also found by \citet{dimauro2016} and 
\citet{dimauro_shear} in the star KIC~4448777. They inferred that the entire core rotates rigidly and provide evidence for an angular velocity gradient around the base of the hydrogen-burning shell. In particular their analysis revealed that the shear layer lies partially inside the hydrogen shell above $r\sim 0.05$~R$_{\rm star}$ and extends across the core-envelope boundary. { KIC~4448777 is not the only giant with a measured gradient in the rotation rate. By comparing the rotation rate of the convective envelope with the measured surface rotation \citep{beck2018} showed that a radial gradient in rotation is also present in  KIC~9163796} 

\begin{figure*}[htb]
\centering
\begin{minipage}{\linewidth}
    \centerline{\includegraphics[width=0.90\linewidth]{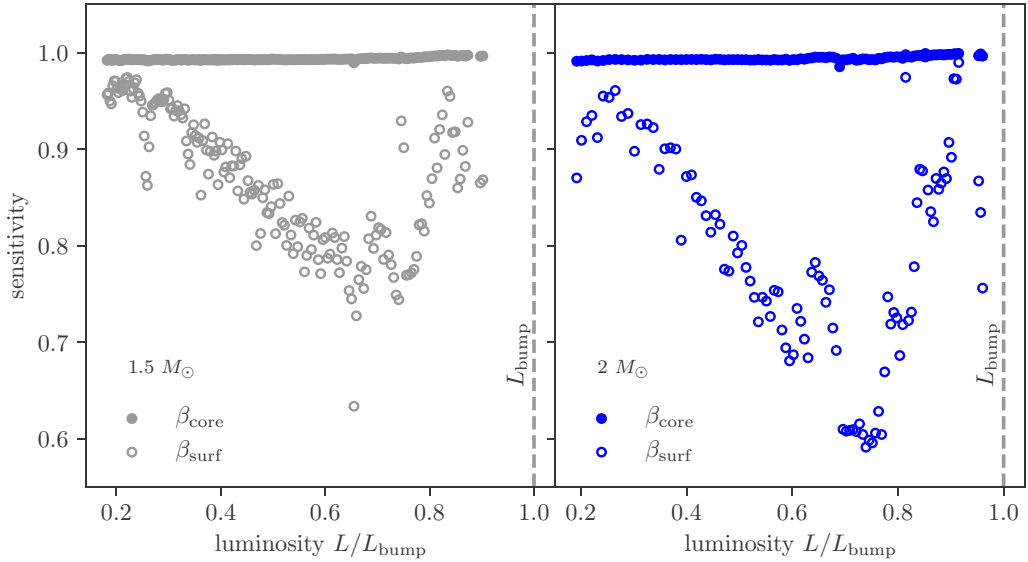}}
\end{minipage}
	\caption{The sensitivity of the resolution, i.e., averaging kernels for models with a mass of 1.5~M$_{\odot}$ (left) and 2.0~M$_{\odot}$ (right) obtained using dipole mode splittings  plotted as a function of luminosity along the red-giant branch. The sensitivity for the core ($\beta_{\rm core}=\int_0^{r_{\rm core}} K(r_0,r)dr $, where $K(r_0, r)$ is the averaging kernel and $r_{\rm core}=0.003$ R$_{\rm star}$) is shown as filled circles, and the surface sensitivity ($\beta_{\rm surf}=1-\int_0^{r_{\rm lim}} K(r_0,r)dr$, $r_{\rm lim}=0.98$ R$_{\rm star}$) is shown as open circles. The vertical dashed lines mark the luminosity of the bump. Note that while there is almost complete sensitivity at the core, the sensitivity to near-surface rotation decreases and then increases again. Image with results from \citet{felix} is courtesy of Felix Ahlborn.
	}
    \label{fig:SENSITIVITY}
\end{figure*}

While inversions have been very successful in determining the rotation rate of red-giant cores, there has been less success when it comes to inversion results of the envelope. Theoretical work in this regards \citep{felix} shows that how sensitive inversion results are to envelope rotation depends on where the star is on the red-giant branch (see Fig.~\ref{fig:SENSITIVITY}). This behaviour is associated with a glitch in the buoyancy frequency which is caused by the composition discontinuity left behind by the convective envelope. The authors only looked at models below the luminosity bump as those are the stars for which mixed modes are observed and rotational splittings are resolved. This theoretical results shows that the sensitivity of the resolution kernels obtained from the inversions imply that we may have better success resolving the envelope of higher-luminosity red giants compared with the lower luminosity ones. 

The most puzzling results about core-rotation have come from ensemble studies of red giants. \citet{mosserspindown} claimed that the cores of red giants, contrary to all expectations, spin down instead of spinning up, as they ascend the red-giant branch. \citet{gehan2018} confirmed this on examining a larger sample of stars, however, they found a smaller trend. \citet{tayar2019} looked at core-helium burning stars rather than ascending branch stars, and they find that the core-rotation rates of these stars decrease strongly with decreasing surface gravity during the core He-burning phase.

The results above clearly indicate that our models of angular momentum transport within stars are deficient. It is clear that the different layers of a star are dynamically coupled and that angular momentum is transferred from the core to the outer layers.  The results have spurred work on trying to understand how angular momentum is transferred between different layers \citep[see e.g.,][etc.]{eggenberger2012, ceillier2013, cantiello2014, belkacem_a, belkacem_b, eggenberger2017, fuller2019}. The processes being studied 
 range from how models of rotation on shells (so-called `shellular rotation') can be modified to give the observed splittings, to meridional circulation, as well as internal gravity modes and magnetic instabilities. 
A comprehensive discussion of the investigations is beyond the scope of this article. Interested readers are directed to \citet{aerts2019} for a review of angular momentum transport in stars.

\vspace{0.75 true cm}

\section{Unsolved issues and neglected data}

The seismic study of red giants is quite new, and their oscillation power spectra, and how best to extract information from them, still poses challenges.

\begin{figure*}
\centering
\begin{minipage}{0.48\linewidth}
    \centerline{\includegraphics[width=\linewidth]{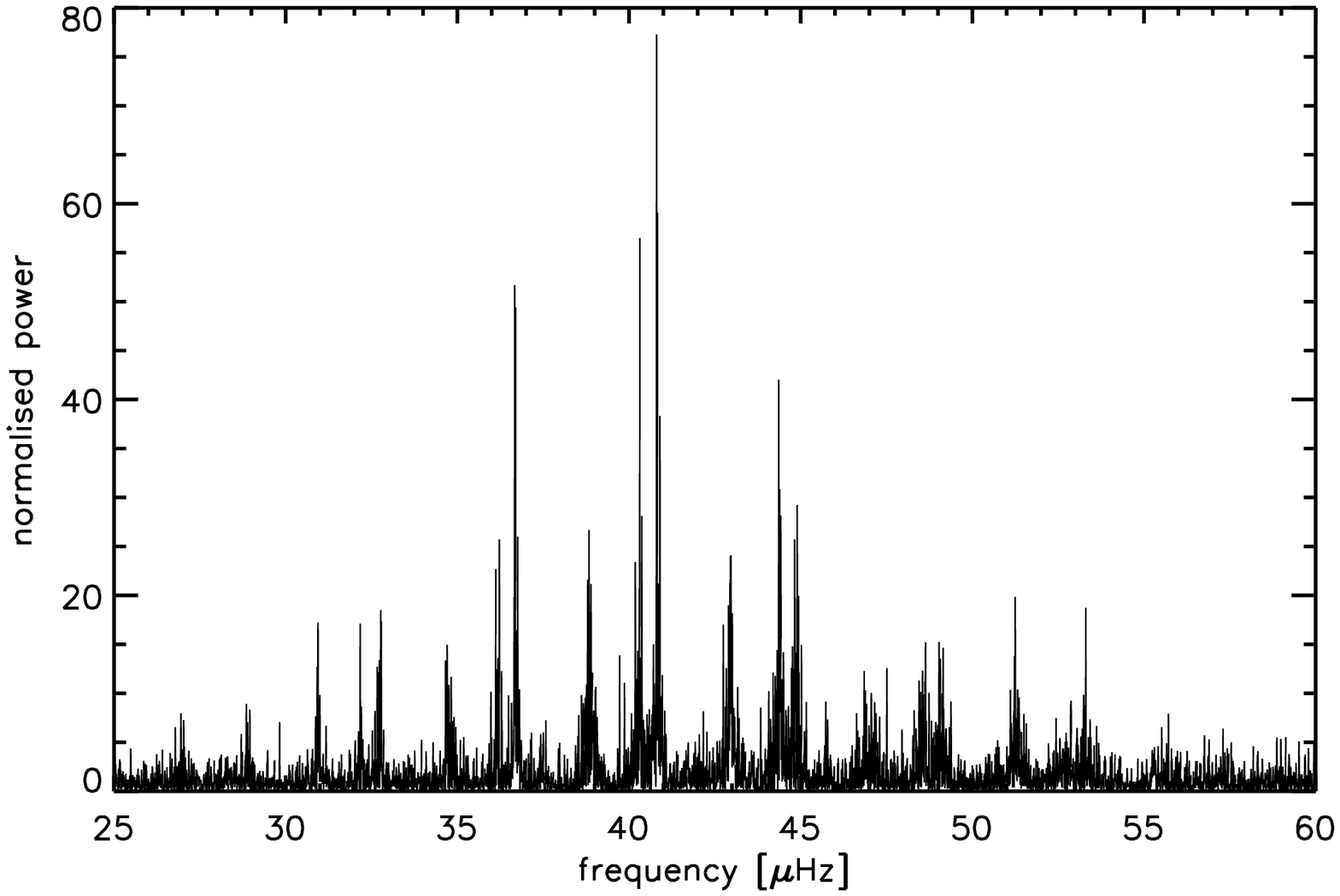}}
\end{minipage}
\begin{minipage}{0.48\linewidth}
    \centerline{\includegraphics[width=\linewidth]{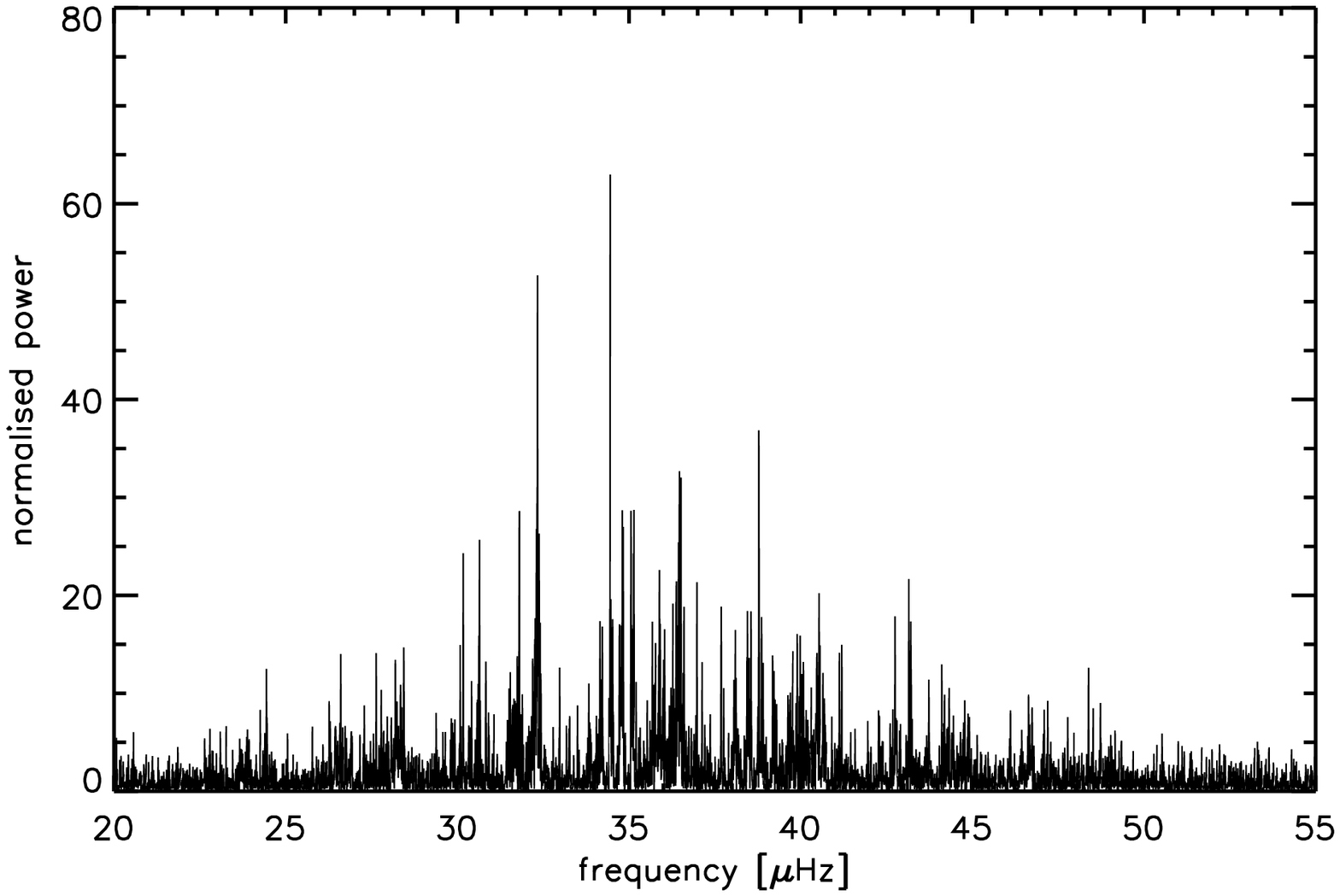}}
\end{minipage}
	\caption{Power spectra of an RGB star (left, KIC 1433730) and an RC star (right, KIC 1161618).} 
   \label{fig:RGB}
 \end{figure*}

It is known empirically that the power spectra of red-clump stars look noisier than those of inert-core red giants, and we can see this in Fig.~\ref{fig:RGB}.  Given that oscillations in these stars are excited by convection, it is to be expected that granulation would play a role in this. Indeed, \citet{savita} showed that the granulation power for red-clump stars can be higher than for a red-giant star of the same radius. However, as expected from theory of stellar convection which states that granulation depends only on \teff, \logg, and metallicity, there does not appear to be any difference between RGB and RC stars when granulation power is plotted against \logg\ \citep[see Figs 7 and 11 in][]{savita}. 

The power spectra of some red giants show suppressed dipole modes \citep{benoitsuppressed, rafasuppressed}. 
 The suppression has been explained as being caused by
remnant magnetic fields in red-giant cores \citep{fuller2015, stello2016}, but this
interpretation has  been challenged \citep{mosser2017}, because the depressed dipole modes
in these red giants are mixed modes and not the pure depressed acoustic modes that would be seen if the modes were suppressed by magnetic fields.  The exact cause of the phenomenon is still under debate and remains a puzzle. See also \citet{loi2017, loi2018} for updated core magnetic field scenarios that could be consistent with the observational findings of suppressed mixed modes by \citet{mosser2017}.

As mentioned earlier, there have not been many attempts to determine the internal structure of many red giants by modelling their frequencies, partly because of issues with correcting for the surface term. However, without modelling the frequencies, a complete analysis of their rotation is not possible either. Thus clearly to understand red giants properly, the available data need to be exploited fully.

Another overlooked piece of information is the signature of glitches in the buoyancy frequency on period spacings.
\citet{cunha2015} presented a theoretical study of how the spacings will be affected. On the observational side, \citet{benoitglitch2015} showed how this signature may be extracted from the data. However, there has been little progress in terms of using the signature to determine the properties of red-giant cores. This is not completely surprising, since interpretation depends both on individual frequencies, which are not commonly extracted, and on models whose physics is uncertain. However, there is a wealth of data which should allow us to make progress despite the uncertainties.

\vspace{0.75 true cm}

\section{Final thoughts}

The oscillation spectra of red giants are fascinating and complicated. They reflect the large density contrast in the interior. The presence of mixed modes has allowed us to study core rotation of these stars, which in turn will allow us to study processes involved with angular-momentum transport inside stars. Methods to exploit mixed modes to study the interior structure of red giants are still under development. Although {\it Kepler} has ceased functioning, the red-giant data obtained by the observatory will be exploited for years to come.

\vspace{0.75 true cm}
\bibliographystyle{frontiersinSCNS_ENG_HUMS} % for Health, Physics and Mathematics articles
\bibliography{main}

\begin{thebibliography}{110}
\providecommand{\natexlab}[1]{#1}
\expandafter\ifx\csname urlstyle\endcsname\relax
  \providecommand{\doi}[1]{doi:\discretionary{}{}{}#1}\else
  \providecommand{\doi}{doi:\discretionary{}{}{}\begingroup
  \urlstyle{rm}\Url}\fi
\providecommand{\selectlanguage}[1]{\relax}
\providecommand{\bibAnnoteFile}[1]{%
  \IfFileExists{#1}{\begin{quotation}\noindent\textsc{Key:} #1\\
  \textsc{Annotation:}\ \input{#1}\end{quotation}}{}}
\providecommand{\bibAnnote}[2]{%
  \begin{quotation}\noindent\textsc{Key:} #1\\
  \textsc{Annotation:}\ #2\end{quotation}}

\bibitem[{{Aerts} et~al.(2010){Aerts}, {Christensen-Dalsgaard}, and
  {Kurtz}}]{aerts}
{Aerts}, C., {Christensen-Dalsgaard}, J., and {Kurtz}, D.~W. (2010).
\newblock \emph{{Asteroseismology}}
\bibAnnoteFile{aerts}

\bibitem[{{Aerts} et~al.(2019){Aerts}, {Mathis}, and {Rogers}}]{aerts2019}
{Aerts}, C., {Mathis}, S., and {Rogers}, T.~M. (2019).
\newblock {Angular Momentum Transport in Stellar Interiors}.
\newblock \emph{\araa} 57, 35--78.
\newblock \doi{10.1146/annurev-astro-091918-104359}
\bibAnnoteFile{aerts2019}

\bibitem[{{Ahlborn} et~al.(2020){Ahlborn}, {Bellinger}, {Hekker}, {Basu}, and
  {Angelou}}]{felix}
{Ahlborn}, F., {Bellinger}, E.~P., {Hekker}, S., {Basu}, S., and {Angelou},
  G.~C. (2020).
\newblock {On the asteroseismic sensitivity to internal rotation along the
  red-giant branch}.
\newblock \emph{arXiv e-prints} , arXiv:2003.08905
\bibAnnoteFile{felix}

\bibitem[{{Aizenman} et~al.(1977){Aizenman}, {Smeyers}, and
  {Weigert}}]{Aizenman1977}
{Aizenman}, M., {Smeyers}, P., and {Weigert}, A. (1977).
\newblock {Avoided Crossing of Modes of Non-radial Stellar Oscillations}.
\newblock \emph{\aap} 58, 41
\bibAnnoteFile{Aizenman1977}

\bibitem[{{Appourchaux} et~al.(2012){Appourchaux}, {Chaplin}, {Garc{\'\i}a},
  {Gruberbauer}, {Verner}, {Antia} et~al.}]{thierry}
{Appourchaux}, T., {Chaplin}, W.~J., {Garc{\'\i}a}, R.~A., {Gruberbauer}, M.,
  {Verner}, G.~A., {Antia}, H.~M., et~al. (2012).
\newblock {Oscillation mode frequencies of 61 main-sequence and subgiant stars
  observed by Kepler}.
\newblock \emph{\aap} 543, A54.
\newblock \doi{10.1051/0004-6361/201218948}
\bibAnnoteFile{thierry}

\bibitem[{{Baglin} et~al.(2006){Baglin}, {Auvergne}, {Boisnard}, {Lam-Trong},
  {Barge}, {Catala} et~al.}]{corot}
{Baglin}, A., {Auvergne}, M., {Boisnard}, L., {Lam-Trong}, T., {Barge}, P.,
  {Catala}, C., et~al. (2006).
\newblock {CoRoT: a high precision photometer for stellar ecolution and
  exoplanet finding}.
\newblock In \emph{36th COSPAR Scientific Assembly}. vol.~36, 3749
\bibAnnoteFile{corot}

\bibitem[{{Ball} and {Gizon}(2014)}]{bg14}
{Ball}, W.~H. and {Gizon}, L. (2014).
\newblock {A new correction of stellar oscillation frequencies for near-surface
  effects}.
\newblock \emph{\aap} 568, A123.
\newblock \doi{10.1051/0004-6361/201424325}
\bibAnnoteFile{bg14}

\bibitem[{{Ball} et~al.(2018){Ball}, {Theme{\ss}l}, and
  {Hekker}}]{balletal2018}
{Ball}, W.~H., {Theme{\ss}l}, N., and {Hekker}, S. (2018).
\newblock {Surface effects on the red giant branch}.
\newblock \emph{\mnras} 478, 4697--4709.
\newblock \doi{10.1093/mnras/sty1141}
\bibAnnoteFile{balletal2018}

\bibitem[{{Balmforth}(1992)}]{balmforth1992}
{Balmforth}, N.~J. (1992).
\newblock {Solar pulsational stability - III. Acoustical excitation by
  turbulent convection}.
\newblock \emph{\mnras} 255, 639.
\newblock \doi{10.1093/mnras/255.4.639}
\bibAnnoteFile{balmforth1992}

\bibitem[{{Basu} and {Chaplin}(2017)}]{mybook}
{Basu}, S. and {Chaplin}, W.~J. (2017).
\newblock \emph{{Asteroseismic Data Analysis: Foundations and Techniques}}
  (Princeton University Press)
\bibAnnoteFile{mybook}

\bibitem[{{Beck} et~al.(2011){Beck}, {Bedding}, {Mosser}, {Stello}, {Garcia},
  {Kallinger} et~al.}]{beck2011}
{Beck}, P.~G., {Bedding}, T.~R., {Mosser}, B., {Stello}, D., {Garcia}, R.~A.,
  {Kallinger}, T., et~al. (2011).
\newblock {Kepler Detected Gravity-Mode Period Spacings in a Red Giant Star}.
\newblock \emph{Science} 332, 205.
\newblock \doi{10.1126/science.1201939}
\bibAnnoteFile{beck2011}

\bibitem[{{Beck} et~al.(2012{\natexlab{a}}){Beck}, {De Ridder}, {Aerts},
  {Kallinger}, {Hekker}, {Garc{\'\i}a} et~al.}]{beck2012a}
{Beck}, P.~G., {De Ridder}, J., {Aerts}, C., {Kallinger}, T., {Hekker}, S.,
  {Garc{\'\i}a}, R.~A., et~al. (2012{\natexlab{a}}).
\newblock {Constraining the core-rotation rate in red-giant stars from Kepler
  space photometry}.
\newblock \emph{Astronomische Nachrichten} 333, 967.
\newblock \doi{10.1002/asna.201211787}
\bibAnnoteFile{beck2012a}

\bibitem[{{Beck} et~al.(2014){Beck}, {Hambleton}, {Vos}, {Kallinger},
  {Bloemen}, {Tkachenko} et~al.}]{beck2014}
{Beck}, P.~G., {Hambleton}, K., {Vos}, J., {Kallinger}, T., {Bloemen}, S.,
  {Tkachenko}, A., et~al. (2014).
\newblock {Pulsating red giant stars in eccentric binary systems discovered
  from Kepler space-based photometry. A sample study and the analysis of KIC
  5006817}.
\newblock \emph{\aap} 564, A36.
\newblock \doi{10.1051/0004-6361/201322477}
\bibAnnoteFile{beck2014}

\bibitem[{{Beck} et~al.(2018){Beck}, {Kallinger}, {Pavlovski}, {Palacios},
  {Tkachenko}, {Mathis} et~al.}]{beck2018}
{Beck}, P.~G., {Kallinger}, T., {Pavlovski}, K., {Palacios}, A., {Tkachenko},
  A., {Mathis}, S., et~al. (2018).
\newblock {Seismic probing of the first dredge-up event through the eccentric
  red-giant and red-giant spectroscopic binary KIC 9163796. How different are
  red-giant stars with a mass ratio of 1.015?}
\newblock \emph{\aap} 612, A22.
\newblock \doi{10.1051/0004-6361/201731269}
\bibAnnoteFile{beck2018}

\bibitem[{{Beck} et~al.(2012{\natexlab{b}}){Beck}, {Montalban}, {Kallinger},
  {De Ridder}, {Aerts}, {Garc{\'\i}a} et~al.}]{beck2012}
{Beck}, P.~G., {Montalban}, J., {Kallinger}, T., {De Ridder}, J., {Aerts}, C.,
  {Garc{\'\i}a}, R.~A., et~al. (2012{\natexlab{b}}).
\newblock {Fast core rotation in red-giant stars as revealed by
  gravity-dominated mixed modes}.
\newblock \emph{\nat} 481, 55--57.
\newblock \doi{10.1038/nature10612}
\bibAnnoteFile{beck2012}

\bibitem[{{Bedding} et~al.(2011{\natexlab{a}}){Bedding}, {Mosser}, {Huber},
  {Montalb{\'a}n}, {Beck}, {Christensen-Dalsgaard} et~al.}]{bedding2011}
{Bedding}, T.~R., {Mosser}, B., {Huber}, D., {Montalb{\'a}n}, J., {Beck}, P.,
  {Christensen-Dalsgaard}, J., et~al. (2011{\natexlab{a}}).
\newblock {Gravity modes as a way to distinguish between hydrogen- and
  helium-burning red giant stars}.
\newblock \emph{\nat} 471, 608--611.
\newblock \doi{10.1038/nature09935}
\bibAnnoteFile{bedding2011}

\bibitem[{{Bedding} et~al.(2011{\natexlab{b}}){Bedding}, {Mosser}, {Huber},
  {Montalb{\'a}n}, {Beck}, {Christensen-Dalsgaard} et~al.}]{bedding}
{Bedding}, T.~R., {Mosser}, B., {Huber}, D., {Montalb{\'a}n}, J., {Beck}, P.,
  {Christensen-Dalsgaard}, J., et~al. (2011{\natexlab{b}}).
\newblock {Gravity modes as a way to distinguish between hydrogen- and
  helium-burning red giant stars}.
\newblock \emph{\nat} 471, 608--611.
\newblock \doi{10.1038/nature09935}
\bibAnnoteFile{bedding}

\bibitem[{{Belkacem} et~al.(2015{\natexlab{a}}){Belkacem}, {Marques}, {Goupil},
  {Mosser}, {Sonoi}, {Ouazzani} et~al.}]{belkacem_b}
{Belkacem}, K., {Marques}, J.~P., {Goupil}, M.~J., {Mosser}, B., {Sonoi}, T.,
  {Ouazzani}, R.~M., et~al. (2015{\natexlab{a}}).
\newblock {Angular momentum redistribution by mixed modes in evolved low-mass
  stars. II. Spin-down of the core of red giants induced by mixed modes}.
\newblock \emph{\aap} 579, A31.
\newblock \doi{10.1051/0004-6361/201526043}
\bibAnnoteFile{belkacem_b}

\bibitem[{{Belkacem} et~al.(2015{\natexlab{b}}){Belkacem}, {Marques}, {Goupil},
  {Sonoi}, {Ouazzani}, {Dupret} et~al.}]{belkacem_a}
{Belkacem}, K., {Marques}, J.~P., {Goupil}, M.~J., {Sonoi}, T., {Ouazzani},
  R.~M., {Dupret}, M.~A., et~al. (2015{\natexlab{b}}).
\newblock {Angular momentum redistribution by mixed modes in evolved low-mass
  stars. I. Theoretical formalism}.
\newblock \emph{\aap} 579, A30.
\newblock \doi{10.1051/0004-6361/201526042}
\bibAnnoteFile{belkacem_a}

\bibitem[{{Borucki} et~al.(2010){Borucki}, {Koch}, {Basri}, {Batalha}, {Brown},
  {Caldwell} et~al.}]{kepler}
{Borucki}, W.~J., {Koch}, D., {Basri}, G., {Batalha}, N., {Brown}, T.,
  {Caldwell}, D., et~al. (2010).
\newblock {Kepler Planet-Detection Mission: Introduction and First Results}.
\newblock \emph{Science} 327, 977.
\newblock \doi{10.1126/science.1185402}
\bibAnnoteFile{kepler}

\bibitem[{{Bovy} et~al.(2014){Bovy}, {Nidever}, {Rix}, {Girardi}, {Zasowski},
  {Chojnowski} et~al.}]{bovy}
{Bovy}, J., {Nidever}, D.~L., {Rix}, H.-W., {Girardi}, L., {Zasowski}, G.,
  {Chojnowski}, S.~D., et~al. (2014).
\newblock {The APOGEE Red-clump Catalog: Precise Distances, Velocities, and
  High-resolution Elemental Abundances over a Large Area of the Milky Way's
  Disk}.
\newblock \emph{\apj} 790, 127.
\newblock \doi{10.1088/0004-637X/790/2/127}
\bibAnnoteFile{bovy}

\bibitem[{{Brown} et~al.(1991){Brown}, {Gilliland}, {Noyes}, and
  {Ramsey}}]{brown1991}
{Brown}, T.~M., {Gilliland}, R.~L., {Noyes}, R.~W., and {Ramsey}, L.~W. (1991).
\newblock {Detection of Possible p-Mode Oscillations on Procyon}.
\newblock \emph{\apj} 368, 599.
\newblock \doi{10.1086/169725}
\bibAnnoteFile{brown1991}

\bibitem[{{Cantiello} et~al.(2014){Cantiello}, {Mankovich}, {Bildsten},
  {Christensen-Dalsgaard}, and {Paxton}}]{cantiello2014}
{Cantiello}, M., {Mankovich}, C., {Bildsten}, L., {Christensen-Dalsgaard}, J.,
  and {Paxton}, B. (2014).
\newblock {Angular Momentum Transport within Evolved Low-mass Stars}.
\newblock \emph{\apj} 788, 93.
\newblock \doi{10.1088/0004-637X/788/1/93}
\bibAnnoteFile{cantiello2014}

\bibitem[{{Ceillier} et~al.(2013){Ceillier}, {Eggenberger}, {Garc{\'\i}a}, and
  {Mathis}}]{ceillier2013}
{Ceillier}, T., {Eggenberger}, P., {Garc{\'\i}a}, R.~A., and {Mathis}, S.
  (2013).
\newblock {Understanding angular momentum transport in red giants: the case of
  KIC 7341231}.
\newblock \emph{\aap} 555, A54.
\newblock \doi{10.1051/0004-6361/201321473}
\bibAnnoteFile{ceillier2013}

\bibitem[{{Chaplin} et~al.(2010){Chaplin}, {Appourchaux}, {Elsworth},
  {Garc{\'\i}a}, {Houdek}, {Karoff} et~al.}]{chaplin2010}
{Chaplin}, W.~J., {Appourchaux}, T., {Elsworth}, Y., {Garc{\'\i}a}, R.~A.,
  {Houdek}, G., {Karoff}, C., et~al. (2010).
\newblock {The Asteroseismic Potential of Kepler: First Results for Solar-Type
  Stars}.
\newblock \emph{\apjl} 713, L169--L175.
\newblock \doi{10.1088/2041-8205/713/2/L169}
\bibAnnoteFile{chaplin2010}

\bibitem[{{Chiappini} et~al.(2015){Chiappini}, {Anders}, {Rodrigues}, {Miglio},
  {Montalb{\'a}n}, {Mosser} et~al.}]{chiappini2015}
{Chiappini}, C., {Anders}, F., {Rodrigues}, T.~S., {Miglio}, A.,
  {Montalb{\'a}n}, J., {Mosser}, B., et~al. (2015).
\newblock {Young [{\ensuremath{\alpha}}/Fe]-enhanced stars discovered by CoRoT
  and APOGEE: What is their origin?}
\newblock \emph{\aap} 576, L12.
\newblock \doi{10.1051/0004-6361/201525865}
\bibAnnoteFile{chiappini2015}

\bibitem[{{Christensen-Dalsgaard}(1988)}]{jcd1988}
{Christensen-Dalsgaard}, J. (1988).
\newblock {A Hertzsprung-Russell Diagram for Stellar Oscillations}.
\newblock In \emph{Advances in Helio- and Asteroseismology}, eds.
  J.~{Christensen-Dalsgaard} and S.~{Frandsen}. vol. 123 of \emph{IAU
  Symposium}, 295
\bibAnnoteFile{jcd1988}

\bibitem[{{Christensen-Dalsgaard} et~al.(2014){Christensen-Dalsgaard}, {Silva
  Aguirre}, {Elsworth}, and {Hekker}}]{JCD2014}
{Christensen-Dalsgaard}, J., {Silva Aguirre}, V., {Elsworth}, Y., and {Hekker},
  S. (2014).
\newblock {On the asymptotic acoustic-mode phase in red giant stars and its
  dependence on evolutionary state}.
\newblock \emph{\mnras} 445, 3685--3693.
\newblock \doi{10.1093/mnras/stu2007}
\bibAnnoteFile{JCD2014}

\bibitem[{{Corsaro} and {De Ridder}(2014)}]{corsaro2014}
{Corsaro}, E. and {De Ridder}, J. (2014).
\newblock {DIAMONDS: A new Bayesian nested sampling tool. Application to peak
  bagging of solar-like oscillations}.
\newblock \emph{\aap} 571, A71.
\newblock \doi{10.1051/0004-6361/201424181}
\bibAnnoteFile{corsaro2014}

\bibitem[{{Cunha} et~al.(2015){Cunha}, {Stello}, {Avelino},
  {Christensen-Dalsgaard}, and {Townsend}}]{cunha2015}
{Cunha}, M.~S., {Stello}, D., {Avelino}, P.~P., {Christensen-Dalsgaard}, J.,
  and {Townsend}, R.~H.~D. (2015).
\newblock {Structural Glitches near the Cores of Red Giants Revealed by
  Oscillations in g-mode Period Spacings from Stellar Models}.
\newblock \emph{\apj} 805, 127.
\newblock \doi{10.1088/0004-637X/805/2/127}
\bibAnnoteFile{cunha2015}

\bibitem[{{Davies} et~al.(2017){Davies}, {Lund}, {Miglio}, {Elsworth},
  {Kuszlewicz}, {North} et~al.}]{guy}
{Davies}, G.~R., {Lund}, M.~N., {Miglio}, A., {Elsworth}, Y., {Kuszlewicz},
  J.~S., {North}, T. S.~H., et~al. (2017).
\newblock {Using red clump stars to correct the Gaia DR1 parallaxes}.
\newblock \emph{\aap} 598, L4.
\newblock \doi{10.1051/0004-6361/201630066}
\bibAnnoteFile{guy}

\bibitem[{{Davies} et~al.(2016){Davies}, {Silva Aguirre}, {Bedding}, {Hand
  berg}, {Lund}, {Chaplin} et~al.}]{kages}
{Davies}, G.~R., {Silva Aguirre}, V., {Bedding}, T.~R., {Hand berg}, R.,
  {Lund}, M.~N., {Chaplin}, W.~J., et~al. (2016).
\newblock {Oscillation frequencies for 35 Kepler solar-type planet-hosting
  stars using Bayesian techniques and machine learning}.
\newblock \emph{\mnras} 456, 2183--2195.
\newblock \doi{10.1093/mnras/stv2593}
\bibAnnoteFile{kages}

\bibitem[{{Deheuvels} et~al.(2014){Deheuvels}, {Do{\u{g}}an}, {Goupil},
  {Appourchaux}, {Benomar}, {Bruntt} et~al.}]{sebastien_six}
{Deheuvels}, S., {Do{\u{g}}an}, G., {Goupil}, M.~J., {Appourchaux}, T.,
  {Benomar}, O., {Bruntt}, H., et~al. (2014).
\newblock {Seismic constraints on the radial dependence of the internal
  rotation profiles of six Kepler subgiants and young red giants}.
\newblock \emph{\aap} 564, A27.
\newblock \doi{10.1051/0004-6361/201322779}
\bibAnnoteFile{sebastien_six}

\bibitem[{{Deheuvels} et~al.(2012){Deheuvels}, {Garc{\'\i}a}, {Chaplin},
  {Basu}, {Antia}, {Appourchaux} et~al.}]{otto}
{Deheuvels}, S., {Garc{\'\i}a}, R.~A., {Chaplin}, W.~J., {Basu}, S., {Antia},
  H.~M., {Appourchaux}, T., et~al. (2012).
\newblock {Seismic Evidence for a Rapidly Rotating Core in a Lower-giant-branch
  Star Observed with Kepler}.
\newblock \emph{\apj} 756, 19.
\newblock \doi{10.1088/0004-637X/756/1/19}
\bibAnnoteFile{otto}

\bibitem[{{Deheuvels} et~al.(2017){Deheuvels}, {Ouazzani}, and
  {Basu}}]{degeneracy}
{Deheuvels}, S., {Ouazzani}, R.~M., and {Basu}, S. (2017).
\newblock {Near-degeneracy effects on the frequencies of rotationally-split
  mixed modes in red giants}.
\newblock \emph{\aap} 605, A75.
\newblock \doi{10.1051/0004-6361/201730786}
\bibAnnoteFile{degeneracy}

\bibitem[{{Di Mauro} et~al.(2016){Di Mauro}, {Ventura}, {Cardini}, {Stello},
  {Christensen-Dalsgaard}, {Dziembowski} et~al.}]{dimauro2016}
{Di Mauro}, M.~P., {Ventura}, R., {Cardini}, D., {Stello}, D.,
  {Christensen-Dalsgaard}, J., {Dziembowski}, W.~A., et~al. (2016).
\newblock {Internal Rotation of the Red-giant Star KIC 4448777 by Means of
  Asteroseismic Inversion}.
\newblock \emph{\apj} 817, 65.
\newblock \doi{10.3847/0004-637X/817/1/65}
\bibAnnoteFile{dimauro2016}

\bibitem[{{Di Mauro} et~al.(2018){Di Mauro}, {Ventura}, {Corsaro}, and {Lustosa
  De Moura}}]{dimauro_shear}
{Di Mauro}, M.~P., {Ventura}, R., {Corsaro}, E., and {Lustosa De Moura}, B.
  (2018).
\newblock {The Rotational Shear Layer inside the Early Red-giant Star KIC
  4448777}.
\newblock \emph{\apj} 862, 9.
\newblock \doi{10.3847/1538-4357/aac7c4}
\bibAnnoteFile{dimauro_shear}

\bibitem[{{Eggenberger} et~al.(2017){Eggenberger}, {Lagarde}, {Miglio},
  {Montalb{\'a}n}, {Ekstr{\"o}m}, {Georgy} et~al.}]{eggenberger2017}
{Eggenberger}, P., {Lagarde}, N., {Miglio}, A., {Montalb{\'a}n}, J.,
  {Ekstr{\"o}m}, S., {Georgy}, C., et~al. (2017).
\newblock {Constraining the efficiency of angular momentum transport with
  asteroseismology of red giants: the effect of stellar mass}.
\newblock \emph{\aap} 599, A18.
\newblock \doi{10.1051/0004-6361/201629459}
\bibAnnoteFile{eggenberger2017}

\bibitem[{{Eggenberger} et~al.(2012){Eggenberger}, {Montalb{\'a}n}, and
  {Miglio}}]{eggenberger2012}
{Eggenberger}, P., {Montalb{\'a}n}, J., and {Miglio}, A. (2012).
\newblock {Angular momentum transport in stellar interiors constrained by
  rotational splittings of mixed modes in red giants}.
\newblock \emph{\aap} 544, L4.
\newblock \doi{10.1051/0004-6361/201219729}
\bibAnnoteFile{eggenberger2012}

\bibitem[{{Elsworth} et~al.(2017){Elsworth}, {Hekker}, {Basu}, and
  {Davies}}]{elsworth2017}
{Elsworth}, Y., {Hekker}, S., {Basu}, S., and {Davies}, G.~R. (2017).
\newblock {A new method for the asteroseismic determination of the evolutionary
  state of red-giant stars}.
\newblock \emph{\mnras} 466, 3344--3352.
\newblock \doi{10.1093/mnras/stw3288}
\bibAnnoteFile{elsworth2017}

\bibitem[{{Elsworth} et~al.(2019){Elsworth}, {Hekker}, {Johnson}, {Kallinger},
  {Mosser}, {Pinsonneault} et~al.}]{elsworth2019}
{Elsworth}, Y., {Hekker}, S., {Johnson}, J.~A., {Kallinger}, T., {Mosser}, B.,
  {Pinsonneault}, M., et~al. (2019).
\newblock {Insights from the APOKASC determination of the evolutionary state of
  red-giant stars by consolidation of different methods}.
\newblock \emph{\mnras} 489, 4641--4657.
\newblock \doi{10.1093/mnras/stz2356}
\bibAnnoteFile{elsworth2019}

\bibitem[{{Epstein} et~al.(2014){Epstein}, {Elsworth}, {Johnson}, {Shetrone},
  {Mosser}, {Hekker} et~al.}]{epstein2014}
{Epstein}, C.~R., {Elsworth}, Y.~P., {Johnson}, J.~A., {Shetrone}, M.,
  {Mosser}, B., {Hekker}, S., et~al. (2014).
\newblock {Testing the Asteroseismic Mass Scale Using Metal-poor Stars
  Characterized with APOGEE and Kepler}.
\newblock \emph{\apjl} 785, L28.
\newblock \doi{10.1088/2041-8205/785/2/L28}
\bibAnnoteFile{epstein2014}

\bibitem[{{Fuller} et~al.(2015){Fuller}, {Cantiello}, {Stello}, {Garcia}, and
  {Bildsten}}]{fuller2015}
{Fuller}, J., {Cantiello}, M., {Stello}, D., {Garcia}, R.~A., and {Bildsten},
  L. (2015).
\newblock {Asteroseismology can reveal strong internal magnetic fields in red
  giant stars}.
\newblock \emph{Science} 350, 423--426.
\newblock \doi{10.1126/science.aac6933}
\bibAnnoteFile{fuller2015}

\bibitem[{{Fuller} et~al.(2019){Fuller}, {Piro}, and {Jermyn}}]{fuller2019}
{Fuller}, J., {Piro}, A.~L., and {Jermyn}, A.~S. (2019).
\newblock {Slowing the spins of stellar cores}.
\newblock \emph{\mnras} 485, 3661--3680.
\newblock \doi{10.1093/mnras/stz514}
\bibAnnoteFile{fuller2019}

\bibitem[{{Garc{\'\i}a} et~al.(2014){Garc{\'\i}a}, {P{\'e}rez Hern{\'a}ndez},
  {Benomar}, {Silva Aguirre}, {Ballot}, {Davies} et~al.}]{rafasuppressed}
{Garc{\'\i}a}, R.~A., {P{\'e}rez Hern{\'a}ndez}, F., {Benomar}, O., {Silva
  Aguirre}, V., {Ballot}, J., {Davies}, G.~R., et~al. (2014).
\newblock {Study of KIC 8561221 observed by Kepler: an early red giant showing
  depressed dipolar modes}.
\newblock \emph{\aap} 563, A84.
\newblock \doi{10.1051/0004-6361/201322823}
\bibAnnoteFile{rafasuppressed}

\bibitem[{{Garc{\'\i}a Saravia Ortiz de Montellano} et~al.(2018){Garc{\'\i}a
  Saravia Ortiz de Montellano}, {Hekker}, and {Theme{\ss}l}}]{garcias2018}
{Garc{\'\i}a Saravia Ortiz de Montellano}, A., {Hekker}, S., and {Theme{\ss}l},
  N. (2018).
\newblock {Automated asteroseismic peak detections}.
\newblock \emph{\mnras} 476, 1470--1496.
\newblock \doi{10.1093/mnras/sty253}
\bibAnnoteFile{garcias2018}

\bibitem[{{Gehan} et~al.(2018){Gehan}, {Mosser}, {Michel}, {Samadi}, and
  {Kallinger}}]{gehan2018}
{Gehan}, C., {Mosser}, B., {Michel}, E., {Samadi}, R., and {Kallinger}, T.
  (2018).
\newblock {Core rotation braking on the red giant branch for various mass
  ranges}.
\newblock \emph{\aap} 616, A24.
\newblock \doi{10.1051/0004-6361/201832822}
\bibAnnoteFile{gehan2018}

\bibitem[{{Girardi}(2016)}]{girardi2016}
{Girardi}, L. (2016).
\newblock {Red Clump Stars}.
\newblock \emph{\araa} 54, 95--133.
\newblock \doi{10.1146/annurev-astro-081915-023354}
\bibAnnoteFile{girardi2016}

\bibitem[{{Gizon} and {Solanki}(2003)}]{gizon2003}
{Gizon}, L. and {Solanki}, S.~K. (2003).
\newblock {Determining the Inclination of the Rotation Axis of a Sun-like
  Star}.
\newblock \emph{\apj} 589, 1009--1019.
\newblock \doi{10.1086/374715}
\bibAnnoteFile{gizon2003}

\bibitem[{{Goldreich} and {Keeley}(1977{\natexlab{a}})}]{goldreich1977a}
{Goldreich}, P. and {Keeley}, D.~A. (1977{\natexlab{a}}).
\newblock {Solar seismology. I. The stability of the solar p-modes.}
\newblock \emph{\apj} 211, 934--942.
\newblock \doi{10.1086/155005}
\bibAnnoteFile{goldreich1977a}

\bibitem[{{Goldreich} and {Keeley}(1977{\natexlab{b}})}]{goldreich1977b}
{Goldreich}, P. and {Keeley}, D.~A. (1977{\natexlab{b}}).
\newblock {Solar seismology. II. The stochastic excitation of the solar p-modes
  by turbulent convection.}
\newblock \emph{\apj} 212, 243--251.
\newblock \doi{10.1086/155043}
\bibAnnoteFile{goldreich1977b}

\bibitem[{{Gough}(1990)}]{dog1990}
{Gough}, D.~O. (1990).
\newblock \emph{{Comments on Helioseismic Inference}}, vol. 367.
\newblock 283.
\newblock \doi{10.1007/3-540-53091-6}
\bibAnnoteFile{dog1990}

\bibitem[{{Goupil} et~al.(2013){Goupil}, {Mosser}, {Marques}, {Ouazzani},
  {Belkacem}, {Lebreton} et~al.}]{goupil2013}
{Goupil}, M.~J., {Mosser}, B., {Marques}, J.~P., {Ouazzani}, R.~M., {Belkacem},
  K., {Lebreton}, Y., et~al. (2013).
\newblock {Seismic diagnostics for transport of angular momentum in stars. II.
  Interpreting observed rotational splittings of slowly rotating red giant
  stars}.
\newblock \emph{\aap} 549, A75.
\newblock \doi{10.1051/0004-6361/201220266}
\bibAnnoteFile{goupil2013}

\bibitem[{{Grundahl} et~al.(2007){Grundahl}, {Kjeldsen},
  {Christensen-Dalsgaard}, {Arentoft}, and {Frandsen}}]{song}
{Grundahl}, F., {Kjeldsen}, H., {Christensen-Dalsgaard}, J., {Arentoft}, T.,
  and {Frandsen}, S. (2007).
\newblock {Stellar Oscillations Network Group}.
\newblock \emph{Communications in Asteroseismology} 150, 300.
\newblock \doi{10.1553/cia150s300}
\bibAnnoteFile{song}

\bibitem[{{Guggenberger} et~al.(2017){Guggenberger}, {Hekker}, {Angelou},
  {Basu}, and {Bellinger}}]{guggenberger2017}
{Guggenberger}, E., {Hekker}, S., {Angelou}, G.~C., {Basu}, S., and
  {Bellinger}, E.~P. (2017).
\newblock {Mitigating the mass dependence in the
  {\ensuremath{\Delta}}{\ensuremath{\nu}} scaling relation of red giant stars}.
\newblock \emph{\mnras} 470, 2069--2078.
\newblock \doi{10.1093/mnras/stx1253}
\bibAnnoteFile{guggenberger2017}

\bibitem[{{Guggenberger} et~al.(2016){Guggenberger}, {Hekker}, {Basu}, and
  {Bellinger}}]{guggenberger2016}
{Guggenberger}, E., {Hekker}, S., {Basu}, S., and {Bellinger}, E. (2016).
\newblock {Significantly improving stellar mass and radius estimates: a new
  reference function for the {\ensuremath{\Delta}}{\ensuremath{\nu}} scaling
  relation}.
\newblock \emph{\mnras} 460, 4277--4281.
\newblock \doi{10.1093/mnras/stw1326}
\bibAnnoteFile{guggenberger2016}

\bibitem[{{Hekker}(2013)}]{hekker2013}
{Hekker}, S. (2013).
\newblock {CoRoT and Kepler results: Solar-like oscillators}.
\newblock \emph{Advances in Space Research} 52, 1581--1592.
\newblock \doi{10.1016/j.asr.2013.08.005}
\bibAnnoteFile{hekker2013}

\bibitem[{{Hekker}(2020)}]{hekker2020}
{Hekker}, S. (2020).
\newblock {Scaling relations for solar-like oscillations: a review}.
\newblock \emph{Frontiers in Astronomy and Space Sciences} 7, 3.
\newblock \doi{10.3389/fspas.2020.00003}
\bibAnnoteFile{hekker2020}

\bibitem[{{Hekker} and {Christensen-Dalsgaard}(2017)}]{hekker2017}
{Hekker}, S. and {Christensen-Dalsgaard}, J. (2017).
\newblock {Giant star seismology}.
\newblock \emph{\aapr} 25, 1.
\newblock \doi{10.1007/s00159-017-0101-x}
\bibAnnoteFile{hekker2017}

\bibitem[{{Hekker} et~al.(2017){Hekker}, {Elsworth}, {Basu}, and
  {Bellinger}}]{hekker2017pipeline}
{Hekker}, S., {Elsworth}, Y., {Basu}, S., and {Bellinger}, E. (2017).
\newblock {Evolutionary states of red-giant stars from grid-based modelling}.
\newblock In \emph{European Physical Journal Web of Conferences}. vol. 160 of
  \emph{European Physical Journal Web of Conferences}, 04006.
\newblock \doi{10.1051/epjconf/201716004006}
\bibAnnoteFile{hekker2017pipeline}

\bibitem[{{Hekker} and {Johnson}(2019)}]{hekker2019}
{Hekker}, S. and {Johnson}, J.~A. (2019).
\newblock {Origin of {\ensuremath{\alpha}}-rich young stars: clues from C, N,
  and O}.
\newblock \emph{\mnras} 487, 4343--4354.
\newblock \doi{10.1093/mnras/stz1554}
\bibAnnoteFile{hekker2019}

\bibitem[{{Hon} et~al.(2018){Hon}, {Stello}, and {Yu}}]{hon}
{Hon}, M., {Stello}, D., and {Yu}, J. (2018).
\newblock {Deep learning classification in asteroseismology using an improved
  neural network: results on 15 000 Kepler red giants and applications to K2
  and TESS data}.
\newblock \emph{\mnras} 476, 3233--3244.
\newblock \doi{10.1093/mnras/sty483}
\bibAnnoteFile{hon}

\bibitem[{{J{\o}rgensen} et~al.(2020){J{\o}rgensen}, {Montalb{\'a}n}, {Miglio},
  {Rendle}, {Davies}, {Buldgen} et~al.}]{newsurface}
{J{\o}rgensen}, A.~S., {Montalb{\'a}n}, J., {Miglio}, A., {Rendle}, B.~M.,
  {Davies}, G.~R., {Buldgen}, G., et~al. (2020).
\newblock {Investigating surface correction relations for RGB stars}.
\newblock \emph{arXiv e-prints} , arXiv:2004.13666
\bibAnnoteFile{newsurface}

\bibitem[{{Kallinger}(2019)}]{kallinger2019}
{Kallinger}, T. (2019).
\newblock {Release note: Massive peak bagging of red giants in the Kepler
  field}.
\newblock \emph{arXiv e-prints} , arXiv:1906.09428
\bibAnnoteFile{kallinger2019}

\bibitem[{{Kallinger} et~al.(2016){Kallinger}, {Hekker}, {Garcia}, {Huber}, and
  {Matthews}}]{kallinger2016}
{Kallinger}, T., {Hekker}, S., {Garcia}, R.~A., {Huber}, D., and {Matthews},
  J.~M. (2016).
\newblock {Precise stellar surface gravities from the time scales of
  convectively driven brightness variations}.
\newblock \emph{Science Advances} 2, 1500654.
\newblock \doi{10.1126/sciadv.1500654}
\bibAnnoteFile{kallinger2016}

\bibitem[{{Kallinger} et~al.(2012){Kallinger}, {Hekker}, {Mosser}, {De Ridder},
  {Bedding}, {Elsworth} et~al.}]{kallinger2012}
{Kallinger}, T., {Hekker}, S., {Mosser}, B., {De Ridder}, J., {Bedding}, T.~R.,
  {Elsworth}, Y.~P., et~al. (2012).
\newblock {Evolutionary influences on the structure of red-giant acoustic
  oscillation spectra from 600d of Kepler observations}.
\newblock \emph{\aap} 541, A51.
\newblock \doi{10.1051/0004-6361/201218854}
\bibAnnoteFile{kallinger2012}

\bibitem[{{Khan} et~al.(2019){Khan}, {Miglio}, {Mosser}, {Arenou}, {Belkacem},
  {Brown} et~al.}]{saniya}
{Khan}, S., {Miglio}, A., {Mosser}, B., {Arenou}, F., {Belkacem}, K., {Brown},
  A.~G.~A., et~al. (2019).
\newblock {New light on the Gaia DR2 parallax zero-point: influence of the
  asteroseismic approach, in and beyond the Kepler field}.
\newblock \emph{\aap} 628, A35.
\newblock \doi{10.1051/0004-6361/201935304}
\bibAnnoteFile{saniya}

\bibitem[{{Kippenhahn} et~al.(2012){Kippenhahn}, {Weigert}, and
  {Weiss}}]{kippenhahn}
{Kippenhahn}, R., {Weigert}, A., and {Weiss}, A. (2012).
\newblock \emph{{Stellar Structure and Evolution}}.
\newblock \doi{10.1007/978-3-642-30304-3}
\bibAnnoteFile{kippenhahn}

\bibitem[{{Kjeldsen} and {Bedding}(1995)}]{KB1995}
{Kjeldsen}, H. and {Bedding}, T.~R. (1995).
\newblock {Amplitudes of stellar oscillations: the implications for
  asteroseismology.}
\newblock \emph{\aap} 293, 87--106
\bibAnnoteFile{KB1995}

\bibitem[{{Kjeldsen} et~al.(2008){Kjeldsen}, {Bedding}, and
  {Christensen-Dalsgaard}}]{kbj}
{Kjeldsen}, H., {Bedding}, T.~R., and {Christensen-Dalsgaard}, J. (2008).
\newblock {Correcting Stellar Oscillation Frequencies for Near-Surface
  Effects}.
\newblock \emph{\apjl} 683, L175.
\newblock \doi{10.1086/591667}
\bibAnnoteFile{kbj}

\bibitem[{{Lian} et~al.(2020){Lian}, {Thomas}, {Maraston}, {Zamora}, {Tayar},
  {Pan} et~al.}]{lian2020}
{Lian}, J., {Thomas}, D., {Maraston}, C., {Zamora}, O., {Tayar}, J., {Pan}, K.,
  et~al. (2020).
\newblock {The age-chemical abundance structure of the Galaxy I: evidence for a
  late-accretion event in the outer disc at z {\ensuremath{\sim}} 0.6}.
\newblock \emph{\mnras} 494, 2561--2575.
\newblock \doi{10.1093/mnras/staa867}
\bibAnnoteFile{lian2020}

\bibitem[{{Loi} and {Papaloizou}(2017)}]{loi2017}
{Loi}, S.~T. and {Papaloizou}, J. C.~B. (2017).
\newblock {Torsional Alfv\'en resonances as an efficient damping mechanism for
  non-radial oscillations in red giant stars}.
\newblock \emph{\mnras} 467, 3212--3225.
\newblock \doi{10.1093/mnras/stx281}
\bibAnnoteFile{loi2017}

\bibitem[{{Loi} and {Papaloizou}(2018)}]{loi2018}
{Loi}, S.~T. and {Papaloizou}, J. C.~B. (2018).
\newblock {Effects of a strong magnetic field on internal gravity waves:
  trapping, phase mixing, reflection, and dynamical chaos}.
\newblock \emph{\mnras} 477, 5338--5357.
\newblock \doi{10.1093/mnras/sty917}
\bibAnnoteFile{loi2018}

\bibitem[{{Lund} et~al.(2017){Lund}, {Silva Aguirre}, {Davies}, {Chaplin},
  {Christensen-Dalsgaard}, {Houdek} et~al.}]{lund}
{Lund}, M.~N., {Silva Aguirre}, V., {Davies}, G.~R., {Chaplin}, W.~J.,
  {Christensen-Dalsgaard}, J., {Houdek}, G., et~al. (2017).
\newblock {Standing on the Shoulders of Dwarfs: the Kepler Asteroseismic LEGACY
  Sample. I. Oscillation Mode Parameters}.
\newblock \emph{\apj} 835, 172.
\newblock \doi{10.3847/1538-4357/835/2/172}
\bibAnnoteFile{lund}

\bibitem[{{Martig} et~al.(2016){Martig}, {Fouesneau}, {Rix}, {Ness},
  {M{\'e}sz{\'a}ros}, {Garc{\'\i}a-Hern{\'a}ndez} et~al.}]{martig2016}
{Martig}, M., {Fouesneau}, M., {Rix}, H.-W., {Ness}, M., {M{\'e}sz{\'a}ros},
  S., {Garc{\'\i}a-Hern{\'a}ndez}, D.~A., et~al. (2016).
\newblock {Red giant masses and ages derived from carbon and nitrogen
  abundances}.
\newblock \emph{\mnras} 456, 3655--3670.
\newblock \doi{10.1093/mnras/stv2830}
\bibAnnoteFile{martig2016}

\bibitem[{{Martig} et~al.(2015){Martig}, {Rix}, {Silva Aguirre}, {Hekker},
  {Mosser}, {Elsworth} et~al.}]{martig2015}
{Martig}, M., {Rix}, H.-W., {Silva Aguirre}, V., {Hekker}, S., {Mosser}, B.,
  {Elsworth}, Y., et~al. (2015).
\newblock {Young {\ensuremath{\alpha}}-enriched giant stars in the solar
  neighbourhood}.
\newblock \emph{\mnras} 451, 2230--2243.
\newblock \doi{10.1093/mnras/stv1071}
\bibAnnoteFile{martig2015}

\bibitem[{{Mathur} et~al.(2016){Mathur}, {Garc{\'\i}a}, {Huber}, {Regulo},
  {Stello}, {Beck} et~al.}]{mathur02016}
{Mathur}, S., {Garc{\'\i}a}, R.~A., {Huber}, D., {Regulo}, C., {Stello}, D.,
  {Beck}, P.~G., et~al. (2016).
\newblock {Probing the Deep End of the Milky Way with Kepler: Asteroseismic
  Analysis of 854 Faint Red Giants Misclassified as Cool Dwarfs}.
\newblock \emph{\apj} 827, 50.
\newblock \doi{10.3847/0004-637X/827/1/50}
\bibAnnoteFile{mathur02016}

\bibitem[{{Mathur} et~al.(2011){Mathur}, {Hekker}, {Trampedach}, {Ballot},
  {Kallinger}, {Buzasi} et~al.}]{savita}
{Mathur}, S., {Hekker}, S., {Trampedach}, R., {Ballot}, J., {Kallinger}, T.,
  {Buzasi}, D., et~al. (2011).
\newblock {Granulation in Red Giants: Observations by the Kepler Mission and
  Three-dimensional Convection Simulations}.
\newblock \emph{\apj} 741, 119.
\newblock \doi{10.1088/0004-637X/741/2/119}
\bibAnnoteFile{savita}

\bibitem[{{McKeever} et~al.(2019){McKeever}, {Basu}, and {Corsaro}}]{mckeever}
{McKeever}, J.~M., {Basu}, S., and {Corsaro}, E. (2019).
\newblock {The Helium Abundance of NGC 6791 from Modeling of Stellar
  Oscillations}.
\newblock \emph{\apj} 874, 180.
\newblock \doi{10.3847/1538-4357/ab0c04}
\bibAnnoteFile{mckeever}

\bibitem[{{Mints} and {Hekker}(2018)}]{mints2018}
{Mints}, A. and {Hekker}, S. (2018).
\newblock {Isochrone fitting in the Gaia era}.
\newblock \emph{\aap} 618, A54.
\newblock \doi{10.1051/0004-6361/201832739}
\bibAnnoteFile{mints2018}

\bibitem[{{Mosser} et~al.(2011){Mosser}, {Barban}, {Montalb{\'a}n}, {Beck},
  {Miglio}, {Belkacem} et~al.}]{mosser2011}
{Mosser}, B., {Barban}, C., {Montalb{\'a}n}, J., {Beck}, P.~G., {Miglio}, A.,
  {Belkacem}, K., et~al. (2011).
\newblock {Mixed modes in red-giant stars observed with CoRoT}.
\newblock \emph{\aap} 532, A86.
\newblock \doi{10.1051/0004-6361/201116825}
\bibAnnoteFile{mosser2011}

\bibitem[{{Mosser} et~al.(2017){Mosser}, {Belkacem}, {Pin{\c{c}}on}, {Takata},
  {Vrard}, {Barban} et~al.}]{mosser2017}
{Mosser}, B., {Belkacem}, K., {Pin{\c{c}}on}, C., {Takata}, M., {Vrard}, M.,
  {Barban}, C., et~al. (2017).
\newblock {Dipole modes with depressed amplitudes in red giants are mixed
  modes}.
\newblock \emph{\aap} 598, A62.
\newblock \doi{10.1051/0004-6361/201629494}
\bibAnnoteFile{mosser2017}

\bibitem[{{Mosser} et~al.(2014){Mosser}, {Benomar}, {Belkacem}, {Goupil},
  {Lagarde}, {Michel} et~al.}]{mosser2014}
{Mosser}, B., {Benomar}, O., {Belkacem}, K., {Goupil}, M.~J., {Lagarde}, N.,
  {Michel}, E., et~al. (2014).
\newblock {Mixed modes in red giants: a window on stellar evolution}.
\newblock \emph{\aap} 572, L5.
\newblock \doi{10.1051/0004-6361/201425039}
\bibAnnoteFile{mosser2014}

\bibitem[{{Mosser} et~al.(2012{\natexlab{a}}){Mosser}, {Elsworth}, {Hekker},
  {Huber}, {Kallinger}, {Mathur} et~al.}]{benoitsuppressed}
{Mosser}, B., {Elsworth}, Y., {Hekker}, S., {Huber}, D., {Kallinger}, T.,
  {Mathur}, S., et~al. (2012{\natexlab{a}}).
\newblock {Characterization of the power excess of solar-like oscillations in
  red giants with Kepler}.
\newblock \emph{\aap} 537, A30.
\newblock \doi{10.1051/0004-6361/201117352}
\bibAnnoteFile{benoitsuppressed}

\bibitem[{{Mosser} et~al.(2012{\natexlab{b}}){Mosser}, {Goupil}, {Belkacem},
  {Marques}, {Beck}, {Bloemen} et~al.}]{mosserspindown}
{Mosser}, B., {Goupil}, M.~J., {Belkacem}, K., {Marques}, J.~P., {Beck}, P.~G.,
  {Bloemen}, S., et~al. (2012{\natexlab{b}}).
\newblock {Spin down of the core rotation in red giants}.
\newblock \emph{\aap} 548, A10.
\newblock \doi{10.1051/0004-6361/201220106}
\bibAnnoteFile{mosserspindown}

\bibitem[{{Mosser} et~al.(2012{\natexlab{c}}){Mosser}, {Goupil}, {Belkacem},
  {Michel}, {Stello}, {Marques} et~al.}]{mosser2012}
{Mosser}, B., {Goupil}, M.~J., {Belkacem}, K., {Michel}, E., {Stello}, D.,
  {Marques}, J.~P., et~al. (2012{\natexlab{c}}).
\newblock {Probing the core structure and evolution of red giants using
  gravity-dominated mixed modes observed with Kepler}.
\newblock \emph{\aap} 540, A143.
\newblock \doi{10.1051/0004-6361/201118519}
\bibAnnoteFile{mosser2012}

\bibitem[{{Mosser} et~al.(2016){Mosser}, {Miglio}, and {CoRot
  Team}}]{mosser2016}
{Mosser}, B., {Miglio}, A., and {CoRot Team} (2016).
\newblock \emph{{IV.2 Pulsating red giant stars}}.
\newblock 197.
\newblock \doi{10.1051/978-2-7598-1876-1.c042}
\bibAnnoteFile{mosser2016}

\bibitem[{{Mosser} et~al.(2015{\natexlab{a}}){Mosser}, {Vrard}, {Belkacem},
  {Deheuvels}, and {Goupil}}]{mosser2015}
{Mosser}, B., {Vrard}, M., {Belkacem}, K., {Deheuvels}, S., and {Goupil}, M.~J.
  (2015{\natexlab{a}}).
\newblock {Period spacings in red giants. I. Disentangling rotation and
  revealing core structure discontinuities}.
\newblock \emph{\aap} 584, A50.
\newblock \doi{10.1051/0004-6361/201527075}
\bibAnnoteFile{mosser2015}

\bibitem[{{Mosser} et~al.(2015{\natexlab{b}}){Mosser}, {Vrard}, {Belkacem},
  {Deheuvels}, and {Goupil}}]{benoitglitch2015}
{Mosser}, B., {Vrard}, M., {Belkacem}, K., {Deheuvels}, S., and {Goupil}, M.~J.
  (2015{\natexlab{b}}).
\newblock {Period spacings in red giants. I. Disentangling rotation and
  revealing core structure discontinuities}.
\newblock \emph{\aap} 584, A50.
\newblock \doi{10.1051/0004-6361/201527075}
\bibAnnoteFile{benoitglitch2015}

\bibitem[{{Nidever} et~al.(2014){Nidever}, {Bovy}, {Bird}, {Andrews}, {Hayden},
  {Holtzman} et~al.}]{nidever2014}
{Nidever}, D.~L., {Bovy}, J., {Bird}, J.~C., {Andrews}, B.~H., {Hayden}, M.,
  {Holtzman}, J., et~al. (2014).
\newblock {Tracing Chemical Evolution over the Extent of the Milky Way's Disk
  with APOGEE Red Clump Stars}.
\newblock \emph{\apj} 796, 38.
\newblock \doi{10.1088/0004-637X/796/1/38}
\bibAnnoteFile{nidever2014}

\bibitem[{{Ong} and {Basu}(2019{\natexlab{a}})}]{ong2019}
{Ong}, J.~M.~J. and {Basu}, S. (2019{\natexlab{a}}).
\newblock {Explaining Deviations from the Scaling Relationship of the Large
  Frequency Separation}.
\newblock \emph{\apj} 870, 41.
\newblock \doi{10.3847/1538-4357/aaf1b5}
\bibAnnoteFile{ong2019}

\bibitem[{{Ong} and {Basu}(2019{\natexlab{b}})}]{epsilon}
{Ong}, J.~M.~J. and {Basu}, S. (2019{\natexlab{b}}).
\newblock {Structural and Evolutionary Diagnostics from Asteroseismic Phase
  Functions}.
\newblock \emph{\apj} 885, 26.
\newblock \doi{10.3847/1538-4357/ab425f}
\bibAnnoteFile{epsilon}

\bibitem[{{Paxton} et~al.(2019){Paxton}, {Smolec}, {Schwab}, {Gautschy},
  {Bildsten}, {Cantiello} et~al.}]{paxton2019}
{Paxton}, B., {Smolec}, R., {Schwab}, J., {Gautschy}, A., {Bildsten}, L.,
  {Cantiello}, M., et~al. (2019).
\newblock {Modules for Experiments in Stellar Astrophysics (MESA): Pulsating
  Variable Stars, Rotation, Convective Boundaries, and Energy Conservation}.
\newblock \emph{\apjs} 243, 10.
\newblock \doi{10.3847/1538-4365/ab2241}
\bibAnnoteFile{paxton2019}

\bibitem[{{Pinsonneault} et~al.(2014){Pinsonneault}, {Elsworth}, {Epstein},
  {Hekker}, {M{\'e}sz{\'a}ros}, {Chaplin} et~al.}]{pinsonneault2014}
{Pinsonneault}, M.~H., {Elsworth}, Y., {Epstein}, C., {Hekker}, S.,
  {M{\'e}sz{\'a}ros}, S., {Chaplin}, W.~J., et~al. (2014).
\newblock {The APOKASC Catalog: An Asteroseismic and Spectroscopic Joint Survey
  of Targets in the Kepler Fields}.
\newblock \emph{\apjs} 215, 19.
\newblock \doi{10.1088/0067-0049/215/2/19}
\bibAnnoteFile{pinsonneault2014}

\bibitem[{{Pinsonneault} et~al.(2018){Pinsonneault}, {Elsworth}, {Tayar},
  {Serenelli}, {Stello}, {Zinn} et~al.}]{pinsonneault2018}
{Pinsonneault}, M.~H., {Elsworth}, Y.~P., {Tayar}, J., {Serenelli}, A.,
  {Stello}, D., {Zinn}, J., et~al. (2018).
\newblock {The Second APOKASC Catalog: The Empirical Approach}.
\newblock \emph{\apjs} 239, 32.
\newblock \doi{10.3847/1538-4365/aaebfd}
\bibAnnoteFile{pinsonneault2018}

\bibitem[{{Salaris} and {Cassisi}(2005)}]{salaris}
{Salaris}, M. and {Cassisi}, S. (2005).
\newblock \emph{{Evolution of Stars and Stellar Populations}}
\bibAnnoteFile{salaris}

\bibitem[{{Schmitt} and {Basu}(2015)}]{joey2015}
{Schmitt}, J.~R. and {Basu}, S. (2015).
\newblock {Modeling the Asteroseismic Surface Term across the HR Diagram}.
\newblock \emph{\apj} 808, 123.
\newblock \doi{10.1088/0004-637X/808/2/123}
\bibAnnoteFile{joey2015}

\bibitem[{{Silva Aguirre} et~al.(2015){Silva Aguirre}, {Davies}, {Basu},
  {Christensen-Dalsgaard}, {Creevey}, {Metcalfe} et~al.}]{sva2015}
{Silva Aguirre}, V., {Davies}, G.~R., {Basu}, S., {Christensen-Dalsgaard}, J.,
  {Creevey}, O., {Metcalfe}, T.~S., et~al. (2015).
\newblock {Ages and fundamental properties of Kepler exoplanet host stars from
  asteroseismology}.
\newblock \emph{\mnras} 452, 2127--2148.
\newblock \doi{10.1093/mnras/stv1388}
\bibAnnoteFile{sva2015}

\bibitem[{{Silva Aguirre} et~al.(2017){Silva Aguirre}, {Lund}, {Antia}, {Ball},
  {Basu}, {Christensen-Dalsgaard} et~al.}]{sva2017}
{Silva Aguirre}, V., {Lund}, M.~N., {Antia}, H.~M., {Ball}, W.~H., {Basu}, S.,
  {Christensen-Dalsgaard}, J., et~al. (2017).
\newblock {Standing on the Shoulders of Dwarfs: the Kepler Asteroseismic LEGACY
  Sample. II.Radii, Masses, and Ages}.
\newblock \emph{\apj} 835, 173.
\newblock \doi{10.3847/1538-4357/835/2/173}
\bibAnnoteFile{sva2017}

\bibitem[{{Sonoi} et~al.(2015){Sonoi}, {Samadi}, {Belkacem}, {Ludwig},
  {Caffau}, and {Mosser}}]{sonoi}
{Sonoi}, T., {Samadi}, R., {Belkacem}, K., {Ludwig}, H.~G., {Caffau}, E., and
  {Mosser}, B. (2015).
\newblock {Surface-effect corrections for solar-like oscillations using 3D
  hydrodynamical simulations. I. Adiabatic oscillations}.
\newblock \emph{\aap} 583, A112.
\newblock \doi{10.1051/0004-6361/201526838}
\bibAnnoteFile{sonoi}

\bibitem[{{Spitoni} et~al.(2020){Spitoni}, {Verma}, {Silva Aguirre}, and
  {Calura}}]{spitoni2020}
{Spitoni}, E., {Verma}, K., {Silva Aguirre}, V., and {Calura}, F. (2020).
\newblock {Galactic archaeology with asteroseismic ages. II. Confirmation of a
  delayed gas infall using Bayesian analysis based on MCMC methods}.
\newblock \emph{\aap} 635, A58.
\newblock \doi{10.1051/0004-6361/201937275}
\bibAnnoteFile{spitoni2020}

\bibitem[{{Stello} et~al.(2016){Stello}, {Cantiello}, {Fuller}, {Huber},
  {Garc{\'\i}a}, {Bedding} et~al.}]{stello2016}
{Stello}, D., {Cantiello}, M., {Fuller}, J., {Huber}, D., {Garc{\'\i}a}, R.~A.,
  {Bedding}, T.~R., et~al. (2016).
\newblock {A prevalence of dynamo-generated magnetic fields in the cores of
  intermediate-mass stars}.
\newblock \emph{\nat} 529, 364--367.
\newblock \doi{10.1038/nature16171}
\bibAnnoteFile{stello2016}

\bibitem[{{Tassoul}(1980)}]{tassoul1980}
{Tassoul}, M. (1980).
\newblock {Asymptotic approximations for stellar nonradial pulsations.}
\newblock \emph{\apjs} 43, 469--490.
\newblock \doi{10.1086/190678}
\bibAnnoteFile{tassoul1980}

\bibitem[{{Tayar} et~al.(2019){Tayar}, {Beck}, {Pinsonneault}, {Garc{\'\i}a},
  and {Mathur}}]{tayar2019}
{Tayar}, J., {Beck}, P.~G., {Pinsonneault}, M.~H., {Garc{\'\i}a}, R.~A., and
  {Mathur}, S. (2019).
\newblock {Core-Envelope Coupling in Intermediate-mass Core-helium Burning
  Stars}.
\newblock \emph{\apj} 887, 203.
\newblock \doi{10.3847/1538-4357/ab558a}
\bibAnnoteFile{tayar2019}

\bibitem[{{Ting} et~al.(2018){Ting}, {Hawkins}, and {Rix}}]{ting}
{Ting}, Y.-S., {Hawkins}, K., and {Rix}, H.-W. (2018).
\newblock {A Large and Pristine Sample of Standard Candles across the Milky
  Way: 100,000 Red Clump Stars with 3\% Contamination}.
\newblock \emph{\apjl} 858, L7.
\newblock \doi{10.3847/2041-8213/aabf8e}
\bibAnnoteFile{ting}

\bibitem[{{Ulrich}(1986)}]{ulrich1986}
{Ulrich}, R.~K. (1986).
\newblock {Determination of Stellar Ages from Asteroseismology}.
\newblock \emph{\apjl} 306, L37.
\newblock \doi{10.1086/184700}
\bibAnnoteFile{ulrich1986}

\bibitem[{{Unno} et~al.(1989){Unno}, {Osaki}, {Ando}, {Saio}, and
  {Shibahashi}}]{unno}
{Unno}, W., {Osaki}, Y., {Ando}, H., {Saio}, H., and {Shibahashi}, H. (1989).
\newblock \emph{{Nonradial oscillations of stars}}
\bibAnnoteFile{unno}

\bibitem[{{Valentini} et~al.(2019){Valentini}, {Chiappini}, {Bossini},
  {Miglio}, {Davies}, {Mosser} et~al.}]{valentini2019}
{Valentini}, M., {Chiappini}, C., {Bossini}, D., {Miglio}, A., {Davies}, G.~R.,
  {Mosser}, B., et~al. (2019).
\newblock {Masses and ages for metal-poor stars. A pilot programme combining
  asteroseismology and high-resolution spectroscopic follow-up of RAVE halo
  stars}.
\newblock \emph{\aap} 627, A173.
\newblock \doi{10.1051/0004-6361/201834081}
\bibAnnoteFile{valentini2019}

\bibitem[{{Vrard} et~al.(2016){Vrard}, {Mosser}, and {Samadi}}]{vrard2016}
{Vrard}, M., {Mosser}, B., and {Samadi}, R. (2016).
\newblock {Period spacings in red giants. II. Automated measurement}.
\newblock \emph{\aap} 588, A87.
\newblock \doi{10.1051/0004-6361/201527259}
\bibAnnoteFile{vrard2016}

\bibitem[{{White} et~al.(2011){White}, {Bedding}, {Stello},
  {Christensen-Dalsgaard}, {Huber}, and {Kjeldsen}}]{whiteetal2011}
{White}, T.~R., {Bedding}, T.~R., {Stello}, D., {Christensen-Dalsgaard}, J.,
  {Huber}, D., and {Kjeldsen}, H. (2011).
\newblock {Calculating Asteroseismic Diagrams for Solar-like Oscillations}.
\newblock \emph{\apj} 743, 161.
\newblock \doi{10.1088/0004-637X/743/2/161}
\bibAnnoteFile{whiteetal2011}

\end{thebibliography}
\vspace{0.75 true cm}

\section*{Appendix A}
Here we present the inlist of the MESA computations used to compute the tracks shown in Fig.~\ref{fig:HRD}. We show here the inlist of the 1~M$_{\odot}$ track and computed the other tracks with exactly the same physics and only changed the mass.
\begin{verbatim}
 ! inlist to evolve a 1 solar mass star

! For the sake of future readers of this file (yourself included),
! ONLY include the controls you are actually using.  DO NOT include
! all of the other controls that simply have their default values.

&star_job

  ! begin with a pre-main sequence model
    create_pre_main_sequence_model = .true.
    
    
  ! begin with saved model
    load_saved_model = .false.

  ! save a model at the end of the run
    save_model_when_terminate = .true.
    save_model_filename = 'bottomAGB10.mod'

  ! display on-screen plots
    pgstar_flag = .true.
 
  pause_before_terminate=.true.

/ !end of star_job namelist


&controls

  ! starting specifications
    initial_mass = 1.0 ! in Msun units

  ! stop when the star nears ZAMS (Lnuc/L > 0.99)
    stop_near_zams = .false.

  ! stop when the center mass fraction of h1 drops below this limit
    xa_central_lower_limit_species(1) = 'he4'
    xa_central_lower_limit(1) = 1d-3
 

   !control output
   history_interval = 1
   write_profiles_flag = .true.
   profile_interval = 20
   profiles_index_name='profile1.0.index'
   profile_data_prefix = 'profile'
   max_num_profile_models = 1000
   star_history_name = 'historym1.0zsun.data'

/ ! end of controls namelist
\end{verbatim}

\end{document}